\documentclass{article}

\PassOptionsToPackage{numbers,compress}{natbib}
\usepackage[main,final]{neurips_2026}

\usepackage[utf8]{inputenc}
\usepackage[T1]{fontenc}
\usepackage{microtype}
\usepackage{graphicx}
\usepackage{subcaption}
\usepackage{booktabs}
\usepackage{amsmath}
\usepackage{amssymb}
\usepackage{mathtools}
\usepackage{amsthm}
\usepackage{multirow}
\usepackage{pifont}
\usepackage{comment}
\usepackage{float}
\usepackage{fvextra}
\usepackage[table,dvipsnames]{xcolor}
\usepackage{url}
\usepackage{hyperref}
\definecolor{linkgreen}{HTML}{006400}
\hypersetup{
  colorlinks=true,
  linkcolor=linkgreen,
  citecolor=linkgreen,
  urlcolor=linkgreen,
  filecolor=linkgreen
}
\usepackage[capitalize,noabbrev]{cleveref}

\usepackage{listings}
\usepackage[most]{tcolorbox}
\tcbuselibrary{listings,skins}
\usepackage{fontawesome5}

\newcommand{\drop}[1]{$_{\scriptstyle\textcolor{red}{\downarrow#1}}$}
\newcommand{\problist}[1]{\tiny\textcolor{darkgray}{\texttt{#1}}}
\newcommand{\good}{\textcolor{green!70!black}{\checkmark}}
\newcommand{\partialgood}{\textcolor{orange}{$\circ$}}
\newcommand{\bad}{\textcolor{red}{\ding{55}}}

\definecolor{cGold}{HTML}{FFF2CC}
\definecolor{cSilver}{HTML}{E7E6E6}
\definecolor{cBronze}{HTML}{F8CBAD}
\definecolor{cBestGroup}{HTML}{DAE8FC}
\definecolor{headergray}{RGB}{240, 240, 240}
\definecolor{verylightgray}{RGB}{243, 245, 248}
\definecolor{easybg}{RGB}{235, 250, 235}
\definecolor{easyframe}{RGB}{100, 180, 100}
\definecolor{medbg}{RGB}{255, 248, 235}
\definecolor{medframe}{RGB}{220, 160, 80}
\definecolor{hardbg}{RGB}{250, 235, 235}
\definecolor{hardframe}{RGB}{200, 100, 100}
\definecolor{cOpen}{HTML}{2E7D32}
\definecolor{cPot}{HTML}{6A1B9A}
\definecolor{cEff}{HTML}{EF6C00}
\definecolor{cBrain}{HTML}{C2185B}
\definecolor{cRobust}{HTML}{1565C0}

\newcommand{\gFirst}[1]{\cellcolor{cGold}#1}
\newcommand{\gSecond}[1]{\cellcolor{cSilver}#1}
\newcommand{\gThird}[1]{\cellcolor{cBronze}#1}
\newcommand{\lBest}[1]{\cellcolor{cBestGroup}#1}
\newcommand{\badgeOpen}{\textcolor{cOpen}{\scriptsize~\faUnlock}}
\newcommand{\badgePot}{\textcolor{cPot}{\scriptsize~\faRocket}}
\newcommand{\badgeEff}{\textcolor{cEff}{\scriptsize~\faBolt}}
\newcommand{\badgeBrain}{\textcolor{cBrain}{\scriptsize~\faBrain}}
\newcommand{\badgeRobust}{\textcolor{cRobust}{\scriptsize~\faBalanceScale}}

\newcolumntype{L}[1]{>{\raggedright\arraybackslash}p{#1}}
\newcommand{\appendixtablestyle}{%
  \small
  \setlength{\tabcolsep}{5pt}%
  \setlength{\abovetopsep}{7pt}%
  \setlength{\aboverulesep}{0pt}%
  \setlength{\belowrulesep}{0pt}%
  \renewcommand{\arraystretch}{1.3}%
}

\newfloat{listing}{tbp}{lol}
\floatname{listing}{Listing}
\lstdefinelanguage{diff}{
  morecomment=[f][\color{red!75!black}]{-},
  morecomment=[f][\color{ForestGreen!80!black}]{+},
  morecomment=[f][\color{blue!70!black}]{@@}
}
\lstdefinestyle{diffstyle}{
  language=diff,
  basicstyle=\small\ttfamily,
  breaklines=true,
  columns=fullflexible,
  keepspaces=true,
  showstringspaces=false,
  numbers=left,
  numbersep=3mm,
  aboveskip=0pt,
  belowskip=0pt
}
\newtcolorbox{listingframe}[1][]{
  colback=gray!5!white,
  colframe=gray!75!black,
  left=5mm,
  enhanced,
  overlay={\begin{tcbclipinterior}\fill[gray!25!white] (frame.south west)
    rectangle ([xshift=5mm]frame.north west);\end{tcbclipinterior}},
  #1
}
\newtcolorbox{promptbox}[1][]{
  colback=blue!5!white,
  colframe=blue!75!black,
  fonttitle=\bfseries,
  title=Prompt,
  enhanced,
  attach boxed title to top left={yshift=-2mm,xshift=5mm},
  boxed title style={colback=blue!75!black},
  #1
}

\newtcblisting{promptboxverb}[1][]{
  listing only,
  listing options={basicstyle=\small\ttfamily,breaklines=true,columns=fullflexible},
  colback=blue!5!white,
  colframe=blue!75!black,
  fonttitle=\bfseries,
  title=Prompt,
  enhanced,
  attach boxed title to top left={yshift=-2mm,xshift=5mm},
  boxed title style={colback=blue!75!black},
  breakable,
  #1
}

\newtcblisting{codebox}[2][]{
  listing options={basicstyle=\small\ttfamily,breaklines=true,columns=fullflexible,numbers=left,numbersep=3mm},
  colback=gray!5!white,
  colframe=gray!75!black,
  listing only,
  left=5mm,
  enhanced,
  overlay={\begin{tcbclipinterior}\fill[gray!25!white] (frame.south west)
    rectangle ([xshift=5mm]frame.north west);\end{tcbclipinterior}},
  #1
}

\newtcblisting{titlecodebox}[3][]{
  listing options={basicstyle=\small\ttfamily,breaklines=true,columns=fullflexible,numbers=left,numbersep=3mm},
  colback=green!5!white,
  colframe=green!65!black,
  fonttitle=\bfseries,
  title=#3,
  listing only,
  left=5mm,
  enhanced,
  attach boxed title to top left={yshift=-2mm,xshift=5mm},
  boxed title style={colback=green!65!black},
  overlay={\begin{tcbclipinterior}\fill[green!20!white] (frame.south west)
    rectangle ([xshift=5mm]frame.north west);\end{tcbclipinterior}},
  #1
}

\theoremstyle{plain}

\theoremstyle{definition}

\theoremstyle{remark}

\makeatletter
\let\neuripssection\section
\let\neuripssubsection\subsection
\renewcommand{\section}{\@startsection{section}{1}{\z@}%
  {-2.0ex \@plus -0.5ex \@minus -0.2ex}%
  {\glueexpr 1.5ex \@plus 0.3ex \@minus 0.2ex - 0.25em\relax}%
  {\large\bf\raggedright}}
\renewcommand{\subsection}{\@startsection{subsection}{2}{\z@}%
  {-1.8ex \@plus -0.5ex \@minus -0.2ex}%
  {\glueexpr 0.8ex \@plus 0.2ex - 0.25em\relax}%
  {\normalsize\bf\raggedright}}
\makeatother
\newcommand{\restorespacing}{%
  \let\section\neuripssection
  \let\subsection\neuripssubsection
  \setlength{\floatsep}{12pt plus 2pt minus 2pt}%
  \setlength{\textfloatsep}{20pt plus 2pt minus 4pt}%
  \setlength{\intextsep}{12pt plus 2pt minus 2pt}}

\title{\textsc{Veri-Sure}: Multi-Agent RTL Code Generation with Temporal Tracing, Slicing and Formal Verification}

\author{%
  Jiale Liu \\
  The University of Edinburgh \\
  \texttt{Jiale.Liu@ed.ac.uk} \\
  \And
  Taiyu Zhou \\
  University of Macau \\
  \texttt{yc57996@um.edu.mo} \\
  \And
  Tianqi Jiang \\
  CUHK, Shenzhen \\
  \texttt{tianqijiang@link.cuhk.edu.cn} \\
}

\begin{document}

\maketitle

\begin{abstract} 
In the rapidly evolving field of Electronic Design Automation (EDA), the deployment of Large Language Models (LLMs) for Register-Transfer Level (RTL) code generation has emerged as a promising direction. However, because RTL code describes concurrent hardware behavior with strict cycle-level timing and thus differs fundamentally from software code, LLM-based RTL code generation remains bottlenecked by (i) sparse feedback and limited test coverage of simulation-centric evaluation, (ii) regressions and repair hallucinations introduced by whole-file regeneration during debugging, and (iii) semantic drift as intent is reinterpreted across repair iterations and agent handoffs. In this work, we propose \textsc{Veri-Sure}, a multi-agent framework that establishes a design contract to align agents' intent and uses a patching mechanism guided by static dependency slicing to perform precise, localized repairs. By integrating a multi-branch verification pipeline that combines trace-driven temporal analysis with formal verification, \textsc{Veri-Sure} yields more robust and reliable RTL code generation. We also introduce \textsc{VerilogEval-v2-EXT}, extending the original benchmark with 53 more industrial-grade design tasks and stratified difficulty levels, and show that \textsc{Veri-Sure} achieves state-of-the-art verified-correct RTL code generation performance, surpassing standalone LLMs and prior agentic systems. Code and dataset are available at \href{https://github.com/xyjoey/Veri-Sure}{GitHub}.

\end{abstract}

\section{Introduction}

The explosive demand for computing power in the era of artificial intelligence and digital transformation continues to push chip designs to billions of transistors, raising the bar for productivity and correctness in Electronic Design Automation (EDA). In practice, the central artifact in the design flow is the Register-Transfer Level (RTL) description, typically written in Hardware Description Languages (HDLs) such as Verilog and SystemVerilog, which specifies clocked state and the combinational logic between registers, precisely capturing concurrent circuit behavior under strict timing constraints and enabling downstream synthesis and implementation. As a result, writing high-quality, synthesizable RTL remains an indispensable yet costly step in modern chip development.

Large language models (LLMs) have recently made rapid progress in software code generation~\citep{jiang_survey_2026, guo_deepseek-coder_2024, liu2026aerogpt, huang_opencoder_2025}, suggesting a compelling opportunity to generate RTL code directly from natural-language specifications. However, RTL code generation poses requirements that are fundamentally different from those of software, since HDL describes concurrent hardware behavior with strict cycle-level timing, protocol semantics, and hierarchical structural constraints~\citep{akyash_simeval_2025}. As a result, general-purpose coding models, which are optimized for autoregressive, sequential text generation, often achieve surface-level syntactic correctness yet fail to reliably preserve temporal intent, handshake protocols, or globally consistent structure~\citep{thakur2024verigen}. These failures are worsened by the scarcity of high-quality HDL data relative to software corpora, which limits the models' exposure to high-quality RTL patterns and actual verification-driven development practices~\citep{liu2024rtlcoder, zhu2025codev}.

To bridge this gap, recent methods have strengthened LLM-based RTL code generation through domain adaptation and tool augmentation, including supervised fine-tuning (SFT)~\citep{thakur2024verigen,lu2024rtllm,liu2024rtlcoder} and reinforcement learning (RL)~\citep{zhu2025codev, min2025improving} on curated HDL corpora, retrieval of reference implementations~\citep{gao2024autovcoder,liu2023chipnemo}, and tool-in-the-loop prompting with simulation feedback~\citep{thakur2023autochip,ho2025verilogcoder,zhao2025mage}. While these techniques improve compile rates and module-level functional pass rates compared to baselines, they remain insufficient for industrial-scale RTL blocks where correctness depends on subtle temporal corner cases and multi-iteration refinements. In particular, we identify that existing LLM-based pipelines are still bottlenecked by (i) sparse feedback and limited test coverage of simulation-centric evaluation, (ii) regression risk, repair hallucinations, and inefficiency caused by whole-file regeneration during debugging, and (iii) semantic drift as cycle-accurate intent is gradually reinterpreted across repair iterations and agent handoffs. Given that hardware bugs can survive simulation and lead to costly tape-out failures, an effective RTL code generation system should not only be able to generate HDL code but also independently perform verification and debugging to ensure the reliability and correctness of the whole design. 

Moreover, progress toward industrial-grade RTL code generation is difficult to measure with existing public benchmarks. Widely used datasets such as VerilogEval-v2~\citep{pinckney_revisiting_2025} are dominated by short, pedagogical modules and under-represent key industrial needs like advanced arithmetic units, communication protocols, buffer/memory control, Clock Domain Crossing (CDC) related patterns, and pipeline-style designs. They also lack difficulty stratification for diagnosing the failure modes of systems. 

To address these challenges, we propose \textsc{Veri-Sure}, a contract-aware multi-agent framework with trace-driven temporal analysis and formal verification. \textsc{Veri-Sure} first converts the natural language specification into a structured design contract that fixes interface semantics and cycle-accurate intent, then uses this contract as the shared source of truth across agents. It then generates candidate RTL code and validates it through simulation. When simulation fails, \textsc{Veri-Sure} performs diagnosis using the temporal analysis and contract-derived formal checks, including assertion checking and Boolean equivalence proofs. Guided by the diagnosis, it applies dependency-slicing-guided patching to perform precise, localized repairs, avoiding unstable whole-file rewrites. Finally, we introduce \textsc{VerilogEval-v2-EXT}, which extends VerilogEval-v2 with 53 industry-grade tasks and a structure-based difficulty level classification to enable finer-grained evaluation. Our main contributions are summarized as follows:

\begin{itemize}
\item We design \textsc{Veri-Sure}, a multi-agent framework for RTL code generation that tightly couples LLM code synthesis with an automated generate-verify-debug loop, leveraging simulation, trace-driven temporal analysis, and formal verification.

\item We propose a dependency-slicing-guided patching mechanism that leverages static dependency slicing over signal relationships to localize failing behaviors and apply precise, minimal edits, avoiding costly whole-file regeneration and reducing regression risks and repair hallucinations.

\item We integrate a multi-branch verification and debugging pipeline that goes beyond simulation by combining trace-driven temporal analysis with formal verification, including assertion-based checking and Boolean equivalence proofs, bringing more reliability and robustness to the debugging loop.

\item We introduce \textsc{VerilogEval-v2-EXT}, extending the original benchmark with expanded task coverage and stratified difficulty levels, and demonstrate state-of-the-art verified-correct RTL code generation performance against standalone LLMs and agentic baselines.
\end{itemize}

\section{Related Work}

\subsection{Automatic RTL Code Generation}

Early studies explored open-loop RTL code generation with general-purpose LLMs via zero-shot or few-shot prompting, and domain-adapted models such as ChipNeMo~\citep{liu2023chipnemo} demonstrated the promise of natural-language-to-RTL workflows. These efforts also highlighted a recurring gap that models may produce syntactically plausible Verilog code while violating cycle-accurate behavior, protocol semantics, or synthesizability constraints.

To improve domain alignment, subsequent work moved beyond vanilla prompting toward domain-adapted training and generation scaffolding. Fine-tuning on curated HDL corpora, as demonstrated by VeriGen~\citep{thakur2024verigen} and RTLLM~\citep{lu2024rtllm}, improves syntax validity and basic module-level functionality. RTLCoder~\citep{liu2024rtlcoder} further filters examples through verification to reduce noisy RTL patterns. BetterV~\citep{pei2024betterv} introduces a generative discriminator to steer outputs toward more verifiable implementations, while CodeV-R1~\citep{zhu2025codev} trains a reasoning model with a distill-then-RL pipeline whose rewards come from testbench-based equivalence checking. Orthogonally, AutoVCoder~\citep{gao2024autovcoder} leverages retrieval-augmented generation to condition synthesis on relevant reference designs.

A complementary line of work strengthens RTL code generation by learning from structured supervision. VeriSeek~\citep{wang2025large} incorporates structure-aware objectives (e.g., abstract syntax tree (AST)-based signals) to encourage hardware-consistent patterns beyond test pass rates, while QiMeng-SALV~\citep{zhang2025qimeng} introduces signal-aware learning to emphasize critical control-signal behavior. VeriLogos~\citep{min2025improving} combines synthetic augmentation with reinforcement learning to improve both diversity and correctness, and DeepRTL~\citep{liu2025deeprtl} explores curriculum-style training to better align natural-language intent with RTL semantics.

Despite steady progress, these methods rely mostly on simulation-centric supervision and lack the iterative repair central to real-world EDA workflows, so they still struggle with concurrent circuit behavior, strict timing requirements, and rare temporal corner cases.

\subsection{Tool-Integrated LLM Systems for HDL}

Recent work has moved toward tool-in-the-loop and agentic pipelines that iteratively generate RTL, run simulation, and repair failures. ChatEDA~\citep{wu2024chateda} showed that an LLM agent can orchestrate EDA tools through their interfaces, AutoChip~\citep{thakur2023autochip} closes the loop using simulator feedback, and RTLFixer~\citep{tsai2024rtlfixer} supports iterative debugging with error-driven edits. These approaches reduce syntax errors and shallow functional bugs. However, their debugging signals are dominated by sparse pass/fail outcomes or textual logs, which provide limited observability for diagnosing timing-dependent failures and localizing root causes in multi-cycle designs.

As tasks scale to larger subsystems and multi-module designs, multi-agent frameworks~\citep{liu2026insightx, liu2026autosurrogate} have been proposed to decompose the workflow into planning, coding, and checking. REvolution~\citep{min2025revolution} frames generation as an evolutionary search with LLM-driven mutations, VerilogCoder~\citep{ho2025verilogcoder} coordinates collaborative agents for generation and validation, and MAGE~\citep{zhao2025mage} improves candidate quality via diversified sampling and targeted debugging. Despite this progress, most agentic systems remain heavily dependent on test coverage, often apply coarse-grained edits that risk regressions, and can accumulate semantic drift across iterations and agent handoffs.

\subsection{Formal Verification and Automated Debugging}

Formal verification aims to mathematically prove the correctness of a design against a specification and is thus also useful for catching bugs in RTL code. Assertion-based verification~\citep{foster2003assertion} embeds temporal properties in the design, which bounded model checking with Boolean satisfiability (SAT) solvers~\citep{biere1999symbolic} can check up to a fixed depth. Since hand-written assertions are costly, AssertLLM~\citep{yan2025assertllm} uses multiple LLMs to generate them from design specifications. For combinational logic, equivalence checking reduces comparing two circuits to proving their miter unsatisfiable~\citep{brand1993verification}, which tools such as SymbiYosys~\citep{yosyshq_symbiyosys} can easily automate with satisfiability modulo theories (SMT) solvers like Z3~\citep{demoura2008z3}.

To achieve automated debugging, prior work has explored both software and hardware contexts. In software, program slicing~\citep{weiser1984program} narrows a failure to the statements that can affect it, and generate-and-validate repair~\citep{legoues2012genprog} searches for patches that pass a test suite. In hardware, CirFix~\citep{ahmad2022cirfix} repairs Verilog by genetic programming and RTL-Repair~\citep{laeufer2024rtl} synthesizes minimal repairs with an SMT solver, but both draw edits from fixed mutation operators or repair templates. LLM-based debuggers lift this limit: MEIC~\citep{xu2024meic} iterates a debug agent with an LLM scoring agent that triggers rollbacks, UVLLM~\citep{hu2025uvllm} couples LLM repair with testing under the Universal Verification Methodology (UVM), and LiK~\citep{yao2025location} localizes bugs from the specification and code alone.

\newpage
\section{The \textsc{Veri-Sure} Framework}

\begin{figure}[htb]
    \centering
    \includegraphics[width=\linewidth]{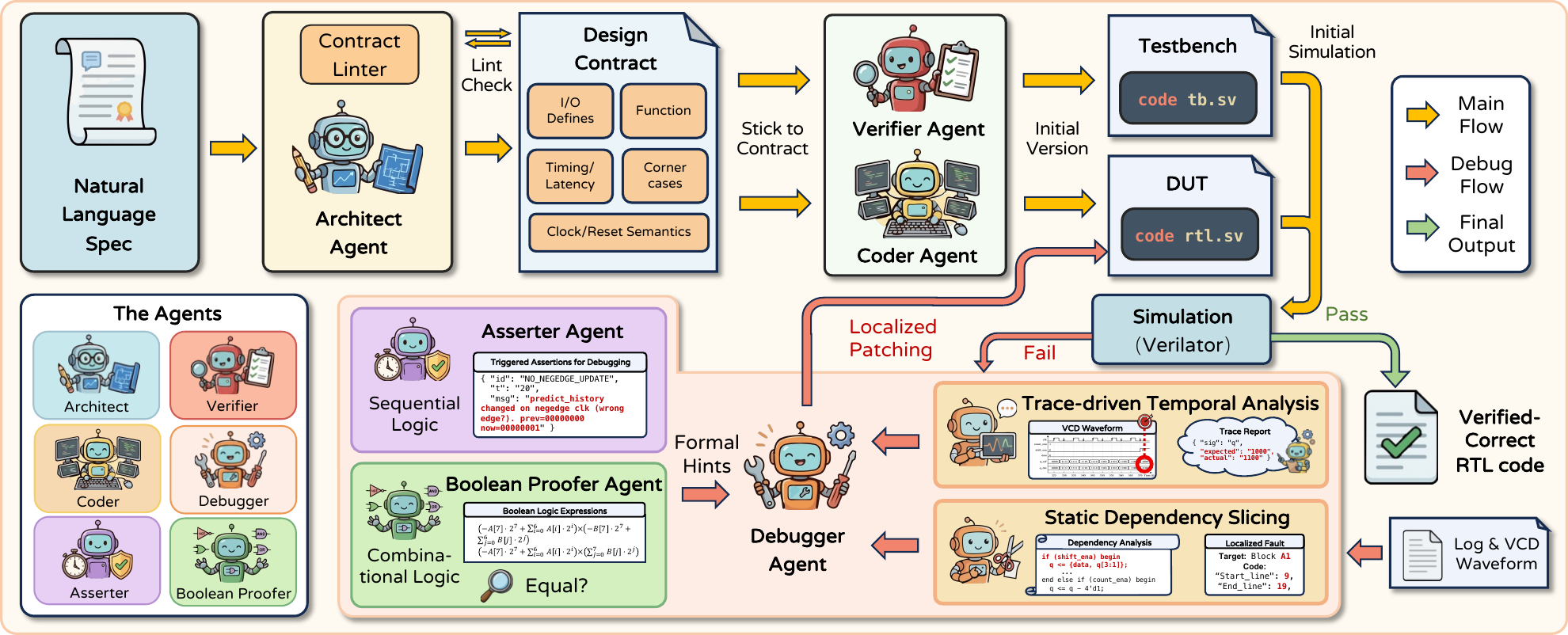}
    \caption{An overview of our \textsc{Veri-Sure} framework.}
    \label{fig:overall-architecture}
    
\end{figure}

We propose \textsc{Veri-Sure}, a multi-agent framework that brings a real EDA-style closed loop for development and testing to RTL synthesis from natural language. Rather than treating generation as a one-shot translation problem, \textsc{Veri-Sure} tightly couples code synthesis with automated validation and actionable debugging signals. Unlike prior multi-agent systems that primarily coordinate through natural-language context and thus risk specification drift and coarse, regression-prone rewrites, \textsc{Veri-Sure} uses trace-driven analysis to localize failures to concrete time windows, signals, and violated requirements, enabling targeted fixes and improving robustness on multi-cycle corner cases. It also integrates simulation with formal checks to continuously enforce design requirements and aid debugging. 

\subsection{Overall Architecture}

\autoref{fig:overall-architecture} shows the workflow of \textsc{Veri-Sure}, a single structured workflow in which agents cooperate by passing typed artifacts rather than free-form messages. The process starts with (1) \textbf{an Architect agent}, which distills the user prompt into a structured JSON-format design contract, capturing the Device Under Test (DUT) interface (ports, clock/reset conventions), key parameters, and cycle-accurate behavioral requirements. When a reference testbench is available, the Architect cross-checks interface and timing details to reduce ambiguity. A linter then validates the contract, and on lint errors, the Architect is re-invoked with the error report. The linted contract is then passed to (2) \textbf{a Verifier agent}, which produces a self-checking testbench whose stimuli and checkers are aligned with the specified requirements, and to (3) \textbf{a Coder agent}, which generates synthesizable RTL code. We then invoke Verilator to compile and simulate the code against the testbench. If all checks pass, the design is accepted.

When the compilation or simulation fails, \textsc{Veri-Sure} enters the debugging loop, in which the failing code is passed to (4) \textbf{a Debugger agent}. Rather than relying solely on textual error messages, the Debugger calls trace-driven temporal analysis and static dependency slicing tools, which turn the raw failure into a trace report of the failing time window and signals, plus a small set of suspect RTL blocks. To provide a denser diagnostic signal, the Debugger also invokes (5) \textbf{an Asserter agent} to generate temporal assertions that expose sequential/protocol violations, and (6) \textbf{a Boolean Proofer agent} to check combinational logic against the contract for counterexamples.  Their outputs are injected into the Debugger's prompt as hints. The Debugger then patches only the suspect blocks, rolling back any patch that does not improve the failure signature, and iterates until all checks pass or the repair budget is exhausted.

\subsection{Contract Design}
\label{sec:contract}

Natural-language RTL specifications are often underspecified in details like reset polarity, sampling edge, or cycle latency, which easily leads to inconsistent interpretations across agents. \textsc{Veri-Sure} mitigates this by having the Architect agent compile the prompt into a compact, structured design contract that makes such choices explicit before any code is written.

The resulting contract records the chosen source of truth, a typed module interface, clock/reset and sequential semantics, per-output latency, a precise functional summary with corner cases, a directed test plan, and guidance for the Verifier, Coder, and Debugger. When the prompt underspecifies details, the Architect makes assumptions explicit so they become checkable requirements. A lightweight contract linter then validates the schema and basic consistency, preventing malformed or contradictory contracts from propagating into code generation and verification.

\subsection{Tracing, Static Slicing \& Patching}
\label{sec:tracing_slicing_patching}

    

\begin{figure}[htbp]
    \centering
    \makebox[\linewidth][s]{%
    \begin{minipage}{0.45\linewidth}
        \centering
        \includegraphics[width=\linewidth]{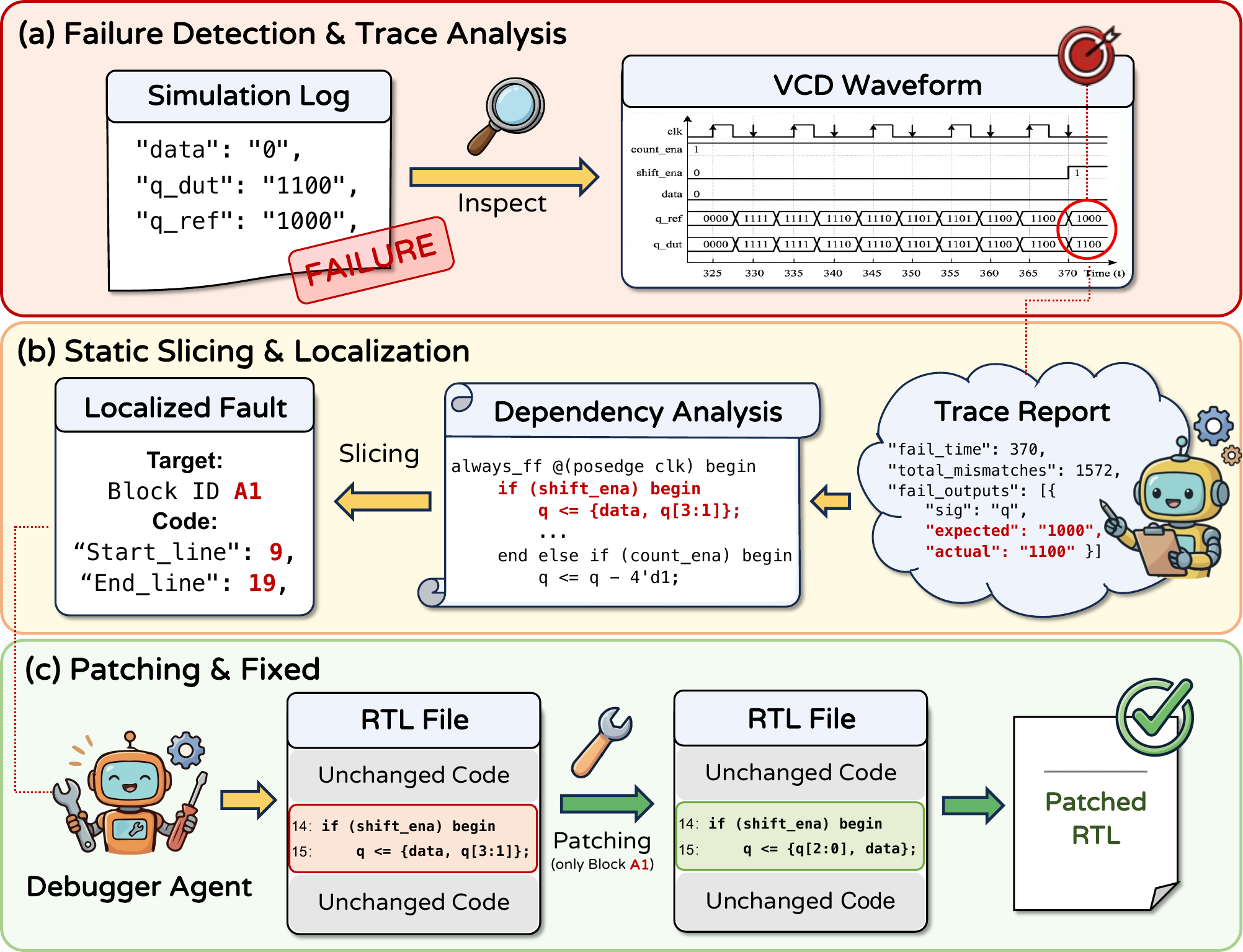}
        \caption{The tracing, static slicing and patching mechanism.}
        \label{fig:verisure-trace-slice-patch}
    \end{minipage}
    \hfill
    \begin{minipage}{0.45\linewidth}
        \centering
        \includegraphics[width=\linewidth]{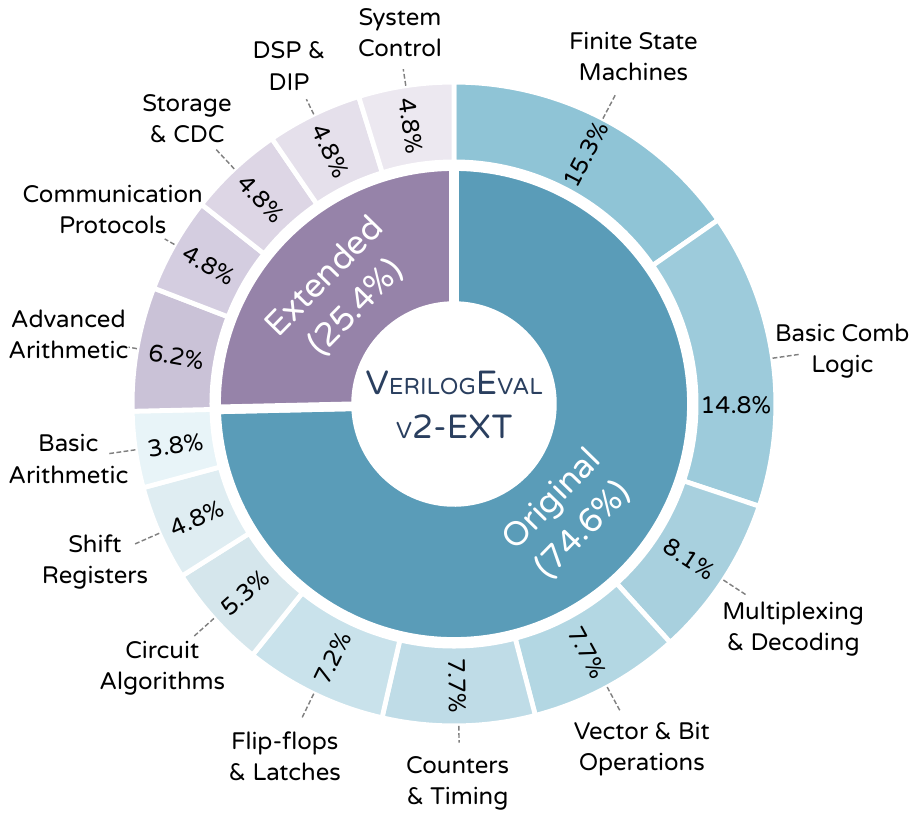}
        \caption{\textsc{VerilogEval-v2-EXT} problem taxonomy.}
        \label{fig:vev2ext_taxonomy}
    \end{minipage}
    }
\end{figure}

When simulation fails, the main challenge is to turn an opaque mismatch count into a minimal, actionable edit without destabilizing unrelated logic. \textsc{Veri-Sure} addresses this with a closed-loop debugging pipeline that couples dynamic evidence (i.e., waveforms) with static structure (i.e., code dependencies), producing a patching task for the Debugger rather than a full-file rewrite, as shown in \autoref{fig:verisure-trace-slice-patch}.

\paragraph{Trace-Driven Temporal Analysis.}
From the simulator log we identify the earliest divergence time and the signals responsible. Let $\mathbf{O}_{\mathrm{DUT}}(t)$ and $\mathbf{O}_{\mathrm{exp}}(t)$ denote the observed and expected output vectors at time $t$, we define:
\begin{equation}
t_f \;=\; \min \{\, t \mid \mathbf{O}_{\mathrm{DUT}}(t) \neq \mathbf{O}_{\mathrm{exp}}(t) \,\}.
\end{equation}
We then extract a short waveform window ending at $t_f$ from the Value Change Dump (VCD) file and sample it on the contract-defined clocking scheme. In addition to raw values, the trace reporter runs a lightweight alignment check to detect systematic off-by-one-cycle or wrong-edge bugs by testing small shifts and reporting the best-matching offset as a timing hint. All findings are summarized into a trace report, including failing signals, the failure cycle, and the relevant trace slice.

\paragraph{Static Dependency Slicing.}
To avoid unfocused edits, the trace slicer restricts attention to the cone of logic that can influence the failing outputs. We parse the RTL into semantic blocks, for example using continuous assignments and always blocks, and compute per-block read/write sets $R(B)$ and $W(B)$. A dependency edge exists when a block reads what another writes:
\begin{equation}
B_j \leftarrow B_i \quad \Leftrightarrow \quad R(B_j)\cap W(B_i)\neq\varnothing.
\end{equation}
Starting from the failing signals as seeds, we perform a bounded backward traversal to obtain a small suspect set $\mathcal{B}_{\mathrm{sus}}$ with block identifiers and line ranges. This converts ``the output mismatched'' into ``these few blocks are the plausible causes,'' dramatically shrinking the search space presented to the Debugger.

\paragraph{Localized Patching.}
The Debugger is only allowed to edit blocks in $\mathcal{B}_{\mathrm{sus}}$ via block-level read/replace operations; the rest of the file is preserved verbatim. Each patch is immediately checked by recompilation and re-simulation, and we apply a simple rollback rule to prevent regressions. Writing $\sigma(\mathrm{RTL})=(t_f,m)$ for the first-failure time and total mismatch count, we keep a patch only if it improves the failure signature, which means later first failure, or fewer mismatches at the same $t_f$; otherwise we revert. This loop grounds the repair process in concrete traces while maintaining stability across iterations.

\subsection{Formal Verification}
\label{sec:formal}

Simulation provides concrete counterexamples but offers limited coverage and often weak and sparse diagnostic signals. \textsc{Veri-Sure} therefore augments the debugging loop with a two-branch, contract-driven verification pipeline: an Asserter for sequential/timing obligations and a Boolean Proofer for combinational equivalence. From the design contract $\widehat{C}$, we derive a small set of checkable obligations:
\begin{equation}
\Phi \;=\; \mathcal{P}(\widehat{C}) \;=\; \Phi_{\mathrm{seq}} \cup \Phi_{\mathrm{comb}},
\end{equation}
which are then discharged by the two verifiers and translated into structured hints for the Debugger.

\paragraph{Asserter for Timing Supervision.}
Using clock/reset semantics and latency annotations in $\widehat{C}$, the Asserter generates a small set of non-intrusive SystemVerilog assertions and attaches them to the DUT via binding. Assertions run under Verilator and report violations with type, implicated signals, and time, effectively diagnosing issues such as wrong-edge sampling, reset mismatches, and off-by-one-cycle latency.

\paragraph{Boolean Proofer for Combinational Equivalence.}
For purely combinational obligations, the Boolean Proofer first infers candidate combinational targets from $\widehat{C}$ (where latency $=0$) and filters out sequentially driven signals via a dependency analysis, avoiding invalid proofs on stateful logic. It then synthesizes a compact reference combinational model from the contract's functional summary and constructs a standard miter that compares DUT and reference under identical inputs. Writing the miter error as
\begin{equation}
e(\mathbf{x}) \;=\; \bigvee_{y \in \mathcal{Y}_{\mathrm{comb}}}\Big(y_{\mathrm{DUT}}(\mathbf{x}) \neq y_{\mathrm{spec}}(\mathbf{x})\Big),
\end{equation}
the goal is to prove $\forall \mathbf{x},\, e(\mathbf{x})=0$ using SymbiYosys. If the proof fails, the resulting counterexample is distilled into a concrete input assignment.

The formal hints produced by the branches are injected into the Debugger prompt, separating timing faults from pure Boolean faults and improving repair generalization. Since these formal artifacts are generated from LLM-derived specifications, we use them only as diagnostic hints and the final acceptance still requires passing the golden testbench. The reliability of these hints is analyzed in \autoref{appendix:formal-signal-reliability}.

\section{The \textsc{VerilogEval-v2-EXT} Benchmark}
\label{sec:vev2ext}

    

We introduce \textsc{VerilogEval-v2-EXT}, an expanded benchmark for evaluating LLMs on realistic RTL code generation. It extends VerilogEval-v2~\citep{pinckney_revisiting_2025} with 53 new, industrial-grade design tasks, increasing the total from 156 to 209 problems. All tasks follow the same evaluation protocol as VerilogEval-v2: each problem provides a natural-language specification, a fully specified module interface, and an executable testbench used for functional checking, enabling direct comparison with prior results.

\begin{figure}[htb]
    \centering
    \setlength{\tabcolsep}{1pt}
    \resizebox{\linewidth}{!}{%
    \begin{tabular}{@{}ccc@{}}
        \includegraphics[width=0.328\linewidth]{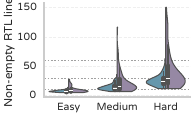} &
        \includegraphics[width=0.328\linewidth]{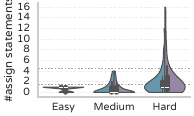} &
        \includegraphics[width=0.328\linewidth]{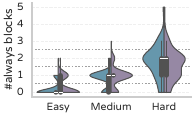} \\
        {\small (a) Code Length} &
        {\small (b) Assign Count} &
        {\small (c) Sequential Structure} \\
        \includegraphics[width=0.328\linewidth]{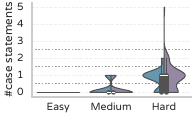} &
        \includegraphics[width=0.328\linewidth]{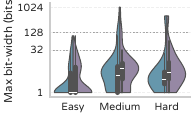} &
        \includegraphics[width=0.328\linewidth]{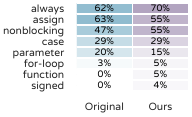} \\
        {\small (d) Control Branching} &
        {\small (e) Data Width} &
        {\small (f) Construct Coverage} \\
    \end{tabular}
    }
    \caption{Dataset complexity statistics comparing VerilogEval-v2 and \textsc{VerilogEval-v2-EXT} (full extended dataset). (a--e) show distributions over problems for different metrics and (f) reports the Verilog construct coverage.}
    \label{fig:vev2ext_stats}
    
\end{figure}

\subsection{Industrial-Grade Extension}
VerilogEval-v2 offers strong coverage of basic digital logic problems, but it under-represents modules that dominate real RTL development (e.g., protocol controllers, buffering, and multi-cycle datapaths). As a result, models may achieve high scores while still failing on engineering-critical behaviors such as latency alignment, corner-case handling, and stateful control logic.

To address these gaps, we curated 53 new problems that reflect common Intellectual Property (IP) blocks and system components, including communication protocols and buffering (e.g., Universal Asynchronous Receiver-Transmitter (UART) interfaces and First-In-First-Out (FIFO) queues), control-dominated designs (e.g., nontrivial Finite State Machines (FSMs)), and richer datapath modules (e.g., advanced arithmetic and DSP-style kernels). \autoref{fig:vev2ext_taxonomy} summarizes the resulting benchmark composition and details are provided in \autoref{appendix:dataset}.

\subsection{Difficulty Annotations}
To support finer-grained analysis beyond an aggregate pass rate, we stratify the 209 problems into Easy/Medium/Hard using a rule-based complexity score derived from observable structural signals, including lines of code, counts of assign/always/case constructs, and datapath width, supplemented with manual review. Full scoring details and thresholds are provided in \autoref{appendix:grading_scheme}. \autoref{fig:vev2ext_stats} shows that the extension increases structural complexity, especially within the Hard subset (e.g., longer solutions with more sequential structure and wider datapaths, up to 1024 bits). Together, these additions make \textsc{VerilogEval-v2-EXT} a more discriminative and industry-aligned testbed for evaluating end-to-end RTL code generation and debugging.

\section{Experiments}

\subsection{Experimental Setup}
\label{sec:exp_setup}

\paragraph{Benchmark.} We evaluate on our \textsc{VerilogEval-v2-EXT} benchmark and report Pass@1 for syntax and functional correctness. Syntax success means the simulator compiles the generated RTL code without errors. Functional success means the compiled DUT passes the provided testbench. Following MAGE~\citep{zhao2025mage}, the golden testbench is provided to the Architect and Verifier as optional context and also serves as the final pass condition. ``Verified-correct'' denotes functional verification in the standard EDA sense, meaning that the design passes the golden testbench.

\paragraph{Baselines.} We compare \textsc{Veri-Sure} against 15 standalone LLMs \citep{openai_introducing_2026,anthropic_introducing_2025,google_deepmind_new_2025,qwen_ai_qwen3-max_2025,qwen_ai_qwen3-coder_2025,mistral_ai_medium_2025,liu_ministral_2026,mistral_ai_introducing_2025,deepseek-ai_deepseek-v32_2025,meta_ai_llama_2025,zhipu_ai_glm-47_2025,team_glm-45_2025,zhang2025qimeng,liu2024rtlcoder,zhu2025codev,min2025improving}, covering both commercial closed-source and recent open-weight models, that generate the DUT in a single attempt. We also include single-agent simulator-feedback baselines that wrap most models with an iterative compile/simulate loop, feeding Verilator logs back for regeneration. Finally, we evaluate representative multi-agent frameworks, MAGE~\citep{zhao2025mage} and VerilogCoder~\citep{ho2025verilogcoder} with the latest backbone model and report all of the comparisons in \autoref{tab:full_comparison}.

\paragraph{Toolchain and Computing.}
All methods use the same testbenches and verification toolchain. We use Verilator~\citep{snyder_verilator} for syntax checking and simulation, chosen for its better support of SystemVerilog assertion-style checks used by our Asserter. For Boolean proof, we use SymbiYosys~\citep{yosyshq_symbiyosys} with a Z3~\citep{demoura2008z3} backend. Commercial models are queried via official APIs; open-weight models are run locally on NVIDIA A100 80GB GPUs. Iterative methods are capped at most $K{=}10$ repair iterations.

\subsection{Main Results}

\begin{table}[htb]
  \centering
  \caption{Performance comparison on \textsc{VerilogEval-v2-EXT} (Pass@1, \%). \textbf{Params} reports total (active) parameters for MoE models, and total for dense models; ``w.'' denotes ``with'' the specified backbone.
  Background colors indicate rankings in each column: \colorbox{cGold}{1st}, \colorbox{cSilver}{2nd}, and \colorbox{cBronze}{3rd} denote the \textbf{Global Performance} across all methods. 
  \colorbox{cBestGroup}{Blue} highlights the \textbf{Group Best} performance if it is not in the global top-3. We assign these badges for standalone LLMs: 
  \badgeOpen~\textbf{Best Open-Weight}: Top-performing open-weight model.
  \badgePot~\textbf{High Potential}: Largest relative gain when enhanced by agents.
  \badgeEff~\textbf{Efficiency King}: Best performance-to-parameter ratio.
  \badgeBrain~\textbf{Reasoning Expert}: Best performance on the ``Hard'' subset.
  \badgeRobust~\textbf{Robust Performer}: Minimal gap between Syntax and Functional scores. 
}
  \label{tab:full_comparison}
  \renewcommand{\arraystretch}{1}
  \resizebox{\linewidth}{!}{%
  \begin{tabular}{@{}lccccccccc@{}}
  \toprule
  \multirow{2}{*}{\textbf{Method}} & \multirow{2}{*}{\textbf{Params}} & \multicolumn{2}{c}{\textbf{Easy} ($n=51$)} & \multicolumn{2}{c}{\textbf{Medium} ($n=91$)} & \multicolumn{2}{c}{\textbf{Hard} ($n=67$)} & \multicolumn{2}{c}{\textbf{Overall}} \\ 
  \cmidrule(lr){3-4} \cmidrule(lr){5-6} \cmidrule(lr){7-8} \cmidrule(lr){9-10}
  & & Syn. & Func. & Syn. & Func. & Syn. & Func. & Syn. & Func. \\ 
  \midrule
  
  \multicolumn{10}{l}{\textit{\textbf{Standalone LLMs}}} \\ 
  \midrule
  GPT-5.2 & - & \gFirst{100.00} & \gThird{94.12} & \gFirst{100.00} & 79.12 & \gFirst{100.00} & 59.70 & \gFirst{100.00} & 76.56 \\
  
  Claude-4.5-Sonnet & - & \gFirst{100.00} & 90.20 & \gFirst{100.00} & 74.73 & \gThird{97.01} & 53.73 & \gThird{99.04} & 71.77 \\
  
  Gemini-3-Pro \badgeBrain & - & \gFirst{100.00} & \gThird{94.12} & \gFirst{100.00} & \lBest{85.71} & \gFirst{100.00} & \lBest{62.69} & \gFirst{100.00} & \lBest{80.38} \\
  
  Qwen3-Max & - & \gThird{96.08} & 86.27 & 93.41 & 54.95 & 86.57 & 37.31 & 91.87 & 56.94 \\
  Mistral-Medium-3.1 & - & \gFirst{100.00} & 78.43 & 87.91 & 52.75 & 77.61 & 19.40 & 87.56 & 48.33 \\
  DeepSeek-3.2 & 685B (37B) & \gThird{96.08} & 80.39 & 96.70 & 64.84 & 85.07 & 41.79 & 92.82 & 61.24 \\
  Qwen3-Coder-Plus & 480B (35B) & 92.16 & 80.39 & 91.21 & 60.44 & 77.61 & 29.85 & 87.08 & 55.50 \\
  LLaMA-4-Maverick & 402B (17B) & \gFirst{100.00} & 86.27 & 85.71 & 53.85 & 76.12 & 32.84 & 86.12 & 55.02 \\
  GLM-4.7 \badgeOpen \badgeRobust & 358B (32B) & \gThird{96.08} & 90.20 & 86.81 & 70.33 & 71.64 & 44.78 & 84.21 & 66.99 \\
  Devstral-2 & 123B & \gSecond{98.04} & 80.39 & 91.21 & 58.24 & 67.16 & 25.37 & 85.17 & 53.11 \\
  Ministral-3-14B \badgePot & 14B & 94.12 & 70.59 & 71.43 & 29.67 & 55.22 & 11.94 & 71.77 & 33.97 \\
  QiMeng-SALV & 7B & \gThird{96.08} & 66.67 & 95.60 & 53.85 & 88.06 & 22.39 & 93.30 & 46.89 \\
  RTL-Coder & 6.7B & 94.12 & 64.71 & 83.52 & 29.67 & 65.67 & 5.97 & 80.38 & 30.62 \\
  CodeV-R1 \badgeEff & 7B & 88.24 & 74.51 & 92.31 & 54.95 & 64.18 & 23.88 & 82.30 & 49.76 \\
  VeriLogos & 7B & 86.27 & 49.02 & 90.11 & 26.37 & 71.64 & 1.49 & 83.25 & 23.92 \\
  
  \midrule
  \multicolumn{10}{l}{\textit{\textbf{Single Agent Systems} (w. Simulator Feedback \& Iterative Fix)}} \\ 
  \midrule
  w. GPT-5.2 & - & \gFirst{100.00} & \gSecond{96.08} & \gFirst{100.00} & 81.32 & \gFirst{100.00} & 59.70 & \gFirst{100.00} & 77.99 \\
  
  w. Gemini-3-Pro & - & \gFirst{100.00} & \gSecond{96.08} & \gFirst{100.00} & \gThird{86.81} & \gFirst{100.00} & \gThird{70.15} & \gFirst{100.00} & \lBest{83.73} \\
  
  w. Qwen3-Max & - & \gFirst{100.00} & \gThird{94.12} & 92.31 & 68.13 & 89.55 & 44.78 & 93.30 & 66.99 \\
  w. Mistral-Medium-3.1 & - & \gSecond{98.04} & 78.43 & 95.60 & 56.04 & 89.55 & 31.34 & 94.26 & 53.59 \\
  
  w. DeepSeek-3.2 & 685B (37B) & \gFirst{100.00} & 84.31 & \gThird{97.80} & 67.03 & 94.03 & 44.78 & 97.13 & 64.11 \\
  
  w. Qwen3-Coder-Plus & 480B (35B) & \gSecond{98.04} & 80.39 & 93.41 & 63.74 & 88.06 & 31.34 & 92.82 & 57.42 \\
  w. LLaMA-4-Maverick & 402B (17B) & \gFirst{100.00} & 92.16 & 91.21 & 59.34 & 79.10 & 32.84 & 89.47 & 58.85 \\
  w. GLM-4.7 & 358B (32B) & \gSecond{98.04} & \gThird{94.12} & 96.70 & 81.32 & 92.54 & 58.21 & 95.69 & 77.03 \\
  w. Devstral-2 & 123B & \gFirst{100.00} & 78.43 & 94.51 & 57.14 & 88.06 & 34.33 & 93.78 & 55.02 \\
  w. Ministral-3-14B & 14B & 94.12 & 76.47 & 79.12 & 48.35 & 55.22 & 19.40 & 75.12 & 45.93 \\
  w. QiMeng-SALV & 7B & \gSecond{98.04} & 74.51 & 94.51 & 51.65 & 88.06 & 17.91 & 93.30 & 46.41 \\
  w. RTL-Coder & 6.7B & 84.31 & 56.86 & 80.22 & 38.46 & 70.15 & 8.96 & 77.99 & 33.49 \\
  w. CodeV-R1 & 7B & \gSecond{98.04} & 84.31 & \gThird{97.80} & 61.54 & 85.07 & 32.84 & 93.78 & 57.89 \\
  w. VeriLogos & 7B & \gThird{96.08} & 56.86 & 86.81 & 24.18 & 82.09 & 4.48 & 87.56 & 25.84 \\
  
  \midrule
  \multicolumn{10}{l}{\textit{\textbf{Multi Agents Systems}}} \\ 
  \midrule
  MAGE (w. GPT-5.2) & - & \gFirst{100.00} & \gSecond{96.08} & \gFirst{100.00} & \gFirst{95.60} & \gSecond{98.51} & \gSecond{77.61} & \gSecond{99.52} & \gSecond{89.95} \\
  
  VerilogCoder (w. GPT-5.2) & - & \gFirst{100.00} & \gThird{94.12} & \gFirst{100.00} & \gSecond{90.11} & \gFirst{100.00} & 68.66 & \gFirst{100.00} & \gThird{84.21} \\
  
  \textsc{Veri-Sure} (w. DS-3.2) & 685B (37B) & \gFirst{100.00} & 90.20 & \gSecond{98.90} & 73.63 & \gThird{97.01} & 53.73 & 98.56 & 71.29 \\
  
  \textbf{\textsc{Veri-Sure} (w. GPT-5.2)} & - & \gFirst{\textbf{100.00}} & \gFirst{\textbf{100.00}} & \gFirst{\textbf{100.00}} & \gFirst{\textbf{95.60}} & \gFirst{\textbf{100.00}} & \gFirst{\textbf{85.07}} & \gFirst{\textbf{100.00}} & \gFirst{\textbf{93.30}} \\
  \bottomrule
  \end{tabular}%
  }
   
\end{table}

\autoref{tab:full_comparison} reports the Pass@1 for syntax and functional correctness on \textsc{VerilogEval-v2-EXT}, further broken down by difficulty. We observe these key things:

\textbf{Frontier closed models are strong out of the box:} Among standalone LLMs, commercial closed-source models deliver consistently high syntax success and strong functional accuracy, especially on Medium and Hard tasks. This suggests that large-scale general pretraining already confers substantial competence in RTL code generation and code-level reasoning, even without hardware-specific fine-tuning.

\textbf{Specialized small models lag behind general large models:} In contrast, smaller models that are fine-tuned for RTL code generation do not necessarily outperform state-of-the-art general-purpose large models. While these models often achieve reasonable syntax rates, their functional pass rates are markedly lower on the harder tasks, indicating that parameter scale and broad pretraining remain critical for multi-step temporal reasoning, corner cases, and control-heavy designs. Practically, this gap is difficult to close by fine-tuning alone because training large backbones is compute-intensive.

\textbf{Feedback improves results, but gains vary by backbone:} Simply feeding simulator logs back to the same model improves syntax and functional accuracy for most backbones, but the magnitude is model-dependent. Stronger models tend to benefit modestly, while weaker or less robust models can gain substantially.

\textbf{Multi-agent approaches are effective and \textsc{Veri-Sure} performs best:} Multi-agent frameworks (MAGE and VerilogCoder, controlled to use GPT-5.2) generally outperform both standalone inference and single-agent feedback, highlighting the value of curated tool use and role specialization.\textsc{Veri-Sure} achieves the best overall functional Pass@1 (93.30\%) with perfect syntax, and the largest advantage on Hard tasks (85.07\% functional). It also boosts the open-weight MoE backbone DeepSeek-3.2 to 71.29\% overall functional Pass@1 (+10.05 pp over standalone; +7.18 pp over single-agent feedback).

\subsection{Ablation Results}

\begin{table}[t!]
  \centering
  \caption{Ablation results of \textsc{Veri-Sure} on \textsc{VerilogEval-v2-EXT} (Func. Pass@1, \%).
  Base model: GPT-5.2. Subscripts denote absolute drop in percentage points vs. the full framework.}
  \label{tab:ablation_singlecol}
  \setlength{\tabcolsep}{3pt}
  \begin{tabular}{@{}p{0.58\columnwidth}cc@{}}
    \toprule
    \textbf{Variant} & \textbf{Hard} & \textbf{Overall} \\
    \midrule
    GPT-5.2 (Standalone) &
      59.70\drop{25.4} & 76.56\drop{16.7} \\
    + Simulator Feedback \& Fix &
      59.70\drop{25.4} & 77.99\drop{15.3} \\
    \midrule
    \rowcolor{gray!20}
    \textsc{Veri-Sure}&
      \textbf{85.07} & \textbf{93.30} \\
    \quad w/o Design Contract &
      82.09\drop{3.0} & 90.43\drop{2.9} \\
    \quad w/o Tracing only &
      76.12\drop{9.0} & 88.04\drop{5.3} \\
    \quad w/o Slicing only &
      77.61\drop{7.5} & 89.00\drop{4.3} \\
    \quad w/o Patching only &
      79.10\drop{6.0} & 89.95\drop{3.4} \\
    \quad w/o Tracing, Slicing \& Patching &
      68.66\drop{16.4} & 82.30\drop{11.0} \\
    \quad w/o Formal Verification &
      73.13\drop{11.9} & 89.47\drop{3.8} \\
    \bottomrule
  \end{tabular}
\end{table}

\autoref{tab:ablation_singlecol} isolates the contribution of each major \textsc{Veri-Sure} component. Overall, \textsc{Veri-Sure} achieves 93.30\% functional pass rate, improving substantially over GPT-5.2 standalone (76.56\%) and the simulator-feedback loop (77.99\%), especially on hard problems (85.07\% vs. 59.70\%). Results also indicate that feedback alone is insufficient for hard RTL tasks, and that most gains come from structured agentic verification and repair.

Generally, removing any major component of \textsc{Veri-Sure} degrades performance. Disabling tracing, slicing, and patching together causes the largest drop, while the finer-grained variants show that all three pieces contribute individually: tracing identifies when and which signals fail, slicing maps those signals to suspect RTL blocks, and patching constrains edits to those blocks. Patching also keeps repairs small: a patch modifies 7.9 lines on average, versus 68.8 for whole-file regeneration (\autoref{appendix:repair-stats}). For Hard problems, Formal Verification is also important: removing it reduces pass rate from 85.07\% to 73.13\% ($-$11.9 pp), indicating that proof-based feedback could help debug deep logical bugs that may not surface under finite simulation.

\subsection{Case Study}
\begin{figure}[tb]
    \centering
    \includegraphics[width=\linewidth]{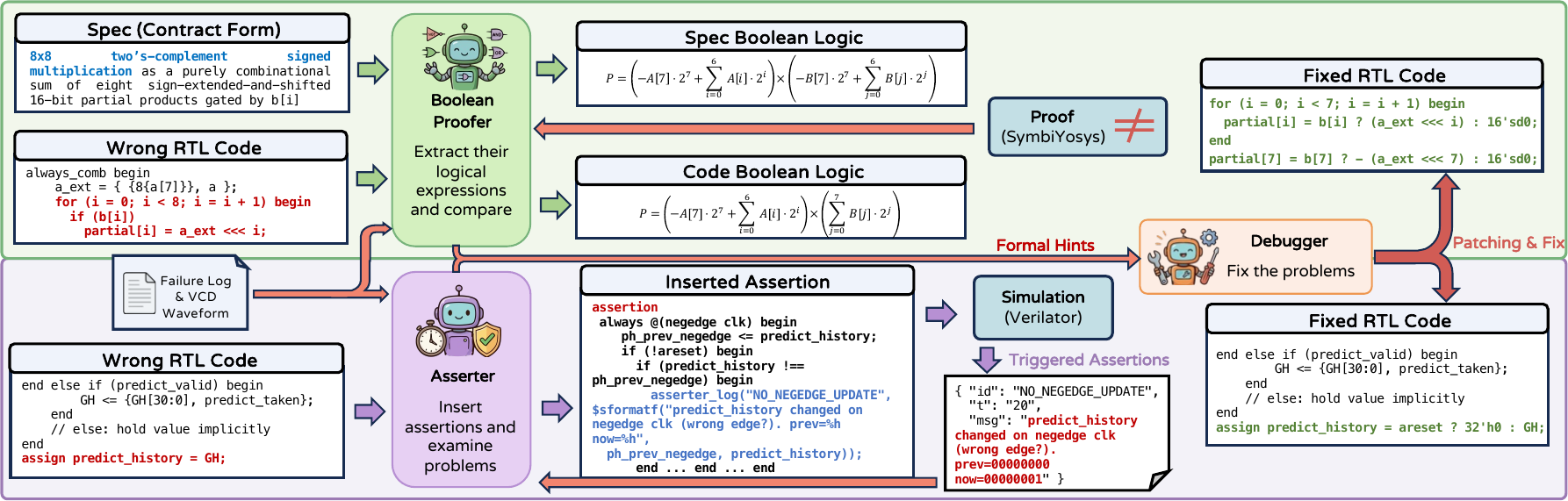}
    \caption{Case study: formal-hint-guided debugging in \textsc{Veri-Sure}. Boolean Proofer and Asserter agents help generate correct RTL code.}
    \label{fig:case_study}
    
\end{figure}

In \autoref{fig:case_study}, we illustrate how \textsc{Veri-Sure} converts failing executions into formal-hint-guided repairs. For an 8$\times$8 two's-complement multiplier (top), the Boolean Proofer derives the contract-level Boolean model and proves inequivalence against the generated RTL via a SymbiYosys miter, pinpointing the missing sign-bit contribution of $b[7]$ and enabling a minimal patch to the partial-product logic. For a control-heavy history-register block (bottom), the Asserter injects a clock/reset assertion that triggers at the first violating cycle and reports the implicated signals/values, guiding the Debugger to fix the edge/reset handling without regenerating the whole module. More detailed success cases can be found in \autoref{appendix:case_study} and we additionally analyze the remaining failures in \autoref{appendix:failure-analysis}.

\section{Failure Analysis}
\label{sec:limitations}

\paragraph{Under-Specified Descriptions.}
Among the 14 tasks that \textsc{Veri-Sure} fails to solve, one is a tooling artifact, and 10 of the remaining 13 true failures are timing-related, such as FSM transitions with ambiguous timing or misaligned output pulses (\autoref{appendix:failure-analysis}). In most of these cases, \textsc{Veri-Sure} locates the faulty logic correctly, but the repair budget runs out before a correct fix is found, usually because the description does not fully determine the intended behavior (e.g., \texttt{Prob117} and \texttt{Prob145}) or the agents fail to infer it. Since patches that do not improve the failure signature are rolled back, extra iterations never make a design worse.

\paragraph{Contract Fidelity.}
We audited all 209 contracts generated by the Architect and found errors in 12 of them (5.7\%), mostly functional contradictions and a few hallucinated constraints, with no interface errors (\autoref{appendix:contract-audit}). Among the 12 tasks, only 3 fail and they still fail without the contract. The other 9 are solved because the Debugger relies on trace reports from simulating the golden testbench, not on the contract. Contracts are also least precise on hard arithmetic, such as floating-point rounding and subnormal handling, which accounts for the remaining 3 (arithmetic) true failures.

\paragraph{Reliability of Formal Hints.}
Since the assertions and Boolean reference models are generated from the LLM-derived contract, they inherit its errors and can also fail on their own. In a full-benchmark run, the Boolean Proofer aborted in 4 of its 30 invocations (SMT solver timeout, unsupported constructs, or an UNKNOWN solver result), and generated assertions can miss the temporal condition behind a failure. Therefore, both branches only provide hints to the Debugger, and acceptance always requires passing the golden testbench.

\section{Conclusion}

In this work, we present \textsc{Veri-Sure}, a contract-aware multi-agent framework that closes the RTL development loop by aligning generation with verification and localized repair. By distilling natural-language intent into a shared design contract, \textsc{Veri-Sure} mitigates semantic drift across agents, while its waveform tracing and static slicing mechanisms enable precise patching that avoids whole-file rewrites and reduces regression risk. Beyond simulation-centric feedback, we integrate formal verification to achieve better debugging capability. To measure the progress on realistic RTL tasks, we also introduce \textsc{VerilogEval-v2-EXT}, including 53 more industrial-grade problems and difficulty stratification. Experiments show that \textsc{Veri-Sure} achieves state-of-the-art verified-correct performance, reaching 93.30\% overall functional pass rate and ablations confirm that the gains come primarily from our debugging pipeline. Overall, \textsc{Veri-Sure} demonstrates that the proposed methods can effectively aid LLMs in generating RTL code, moving automated design closer toward silicon-grade correctness.

\clearpage
\restorespacing

\begin{ack}
We thank The University of Edinburgh's ELM service and EDINA team for supporting the API access. This work also used the Isambard-AI National AI Research Resource (AIRR), operated by the University of Bristol and funded by the UK Government's Department for Science, Innovation and Technology (DSIT) via UK Research and Innovation, and by the Science and Technology Facilities Council [ST/AIRR/I-A-I/1023].
\end{ack}

\bibliographystyle{plainnat}
\bibliography{example_paper}

\newpage
\appendix

\section*{Appendices}

We supplement some additional technical content here in the appendices. Specifically, \autoref{appendix:dataset} presents additional details of the
\textsc{VerilogEval-v2-EXT} benchmark; \autoref{appendix:verisure_impl} documents
implementation details of the \textsc{Veri-Sure} framework; \autoref{appendix:experiments} provides failure, design-contract, formal-signal reliability, repairing mechanism, and cost analyses; \autoref{appendix:case_study}
reports three successful case studies; and \autoref{appendix:framework_comparison} offers a
detailed comparison with other related frameworks.

\section{The \textsc{VerilogEval-v2-EXT} Benchmark}
\label{appendix:dataset}

This appendix provides additional details for \textsc{VerilogEval-v2-EXT}, including (i) a full taxonomy and coverage analysis of the original VerilogEval-v2 tasks, (ii) the motivations and design targets of our 53 newly added industrial-grade tasks, and (iii) the rule-based difficulty grading scheme used in the paper.

\subsection{Summary of Existing Problems}

The original VerilogEval-v2 benchmark contains 156 RTL code generation tasks. While it offers strong coverage of basic digital logic problems, we find it under-represents the types of modules that dominate real RTL development, such as protocol controllers, buffering, and multi-cycle datapaths. As a result, models can achieve high scores by mastering local syntax patterns and short combinational reasoning, while still failing on engineering-critical behaviors like latency alignment, corner-case handling, and stateful control logic. To make this limitation explicit and to identify concrete coverage gaps, we manually reviewed all 156 problems and grouped them into nine categories, summarized in \autoref{tab:vev2_taxonomy_original}.

\paragraph{Taxonomy Methodology.} Each problem is assigned to the category that most strongly determines its implementation structure (e.g., whether the solution is dominated by combinational datapath wiring, sequential state machines, or algorithmic transformations). When a problem touches multiple concepts, we classify it by the dominant design pattern required to pass the golden testbench.

\begin{table}[h!]
  \centering
  \caption{Full taxonomy of the original VerilogEval-v2 benchmark. We group problems into nine categories to reveal coverage biases and industrial blind spots.}
  \label{tab:vev2_taxonomy_original}
  \appendixtablestyle\footnotesize
  \rowcolors{2}{verylightgray}{white}
  \begin{tabular}{@{}L{0.18\textwidth} c L{0.38\textwidth} L{0.32\textwidth}@{}}
    \toprule
    \textbf{Category} & \textbf{\#} & \textbf{Problem IDs} & \textbf{Description \& Limitations} \\ 
    \midrule
    
    I. Basic Logic Gates & 16 & 
    \problist{Prob001, Prob005, Prob007, Prob011, Prob012, Prob013, Prob014, Prob019, Prob026, Prob043, Prob051, Prob052, Prob059, Prob077, Prob087, Prob092} & 
    Gate-level/bit-level Boolean logic. Limited depth and limited temporal reasoning. \\

    II. Vectors \& Bit Manipulation & 14 & 
    \problist{Prob004, Prob006, Prob015, Prob023, Prob025, Prob032, Prob042, Prob044, Prob055, Prob062, Prob064, Prob070, Prob094, Prob097} & 
    Concatenation, slicing, bit-reversal, and parity. Does not cover real numeric formats (fixed-/floating-point). \\

    III. Mux \& Decoding & 11 & 
    \problist{Prob017, Prob018, Prob021, Prob022, Prob039, Prob071, Prob076, Prob093, Prob104, Prob112, Prob114} & 
    Standard mux/decoder/encoder usage. Mostly textbook patterns; little parameterization. \\

    IV. Basic Arithmetic & 9 & 
    \problist{Prob009, Prob016, Prob024, Prob027, Prob030, Prob033, Prob065, Prob081, Prob123} & 
    Mostly add/sub/popcount. Notably lacks multipliers/dividers, saturation, and complex datapaths. \\

    V. Latches \& Flip-Flops & 14 & 
    \problist{Prob002, Prob003, Prob008, Prob028, Prob029, Prob031, Prob034, Prob041, Prob046, Prob047, Prob048, Prob049, Prob053, Prob073} & 
    Basic sequential primitives (DFF/latch). Almost entirely single-clock; does not test CDC patterns. \\

    VI. Counters, Timers \& Edge Det. & 15 & 
    \problist{Prob035, Prob037, Prob038, Prob040, Prob045, Prob054, Prob066, Prob067, Prob068, Prob075, Prob078, Prob080, Prob141, Prob151, Prob156} & 
    Counting/timing logic. Typically linear control, limited handshake or flow-control complexity. \\

    VII. Shift Registers & 12 & 
    \problist{Prob060, Prob061, Prob063, Prob082, Prob084, Prob085, Prob086, Prob095, Prob096, Prob105, Prob115, Prob118} & 
    Shifting/LFSR/pattern detection. Does not include streaming buffers (FIFO) or backpressure. \\

    VIII. Finite State Machines (FSMs) & 42 & 
    \problist{Prob036, Prob056, Prob072, Prob074, Prob079, Prob088, Prob089, Prob091, Prob099, Prob100, Prob106, Prob107, Prob108, Prob109, Prob110, Prob111, Prob119, Prob120, Prob121, Prob124, Prob127, Prob128, Prob129, Prob133, Prob134, Prob135, Prob136, Prob137, Prob138, Prob139, Prob140, Prob142, Prob143, Prob144, Prob146, Prob147, Prob148, Prob149, Prob150, Prob152, Prob154, Prob155} & 
    Largest category. Many are variants of tutorial sequence detectors; few reflect on-chip protocols.\\

    IX. Interpretation \& Algorithms & 23 & 
    \problist{Prob010, Prob020, Prob050, Prob057, Prob058, Prob069, Prob083, Prob090, Prob098, Prob101, Prob102, Prob103, Prob113, Prob116, Prob117, Prob122, Prob125, Prob126, Prob130, Prob131, Prob132, Prob145, Prob153} & 
    Algorithmic ``translation'' tasks (e.g., Game of Life, waveform interpretation). Less representative of microarchitecture. \\
    \bottomrule
  \end{tabular}
\end{table}

\paragraph{Key Observations and Limitations.}
This taxonomy highlights why simulation-only pass rates on VerilogEval-v2 can overestimate progress toward silicon-grade RTL code generation:
\begin{itemize}
  \item \textbf{Over-emphasis on short, textbook patterns:} A large fraction of tasks are small-scale logic exercises (Categories I--III) or repetitive FSM variants (Category VIII).
  \item \textbf{Severe under-coverage of industrial datapaths:} ``Basic Arithmetic'' comprises only 9/156 tasks, largely limited to simple integer operations.
  \item \textbf{Missing protocol and buffering semantics:} The original set does not systematically test standard communication protocols (UART/SPI/I\textsuperscript{2}C) or ubiquitous flow-control primitives (FIFOs, skid buffers).
  \item \textbf{Limited exposure to CDC and multi-clock assumptions:} Most tasks assume a single clock domain, ignoring reset-domain reasoning and CDC safety.
\end{itemize}

\subsection{New Problems}

To address the coverage gaps above, we introduce \textsc{VerilogEval-v2-EXT} by adding 53 new industrial-grade tasks, bringing the total to 209 problems. The new problems are designed to be industry-aligned, specifically targeting multi-cycle behaviors and synthesizable constructs. We aim to cover the missing areas detailed in \autoref{tab:vev2ext_new_categories}.

\begin{table}[h!]
  \centering
  \caption{Categories of the 53 new tasks in \textsc{VerilogEval-v2-EXT}. The categories are chosen to reflect common IP/SoC RTL development patterns that are sparse or absent in the original VerilogEval-v2 benchmark.}
  \label{tab:vev2ext_new_categories}
  \appendixtablestyle
  \rowcolors{2}{verylightgray}{white}
  \begin{tabular}{@{}L{0.21\textwidth} L{0.35\textwidth} L{0.37\textwidth}@{}}
    \toprule
    \textbf{Category} & \textbf{Motivation} & \textbf{Representative RTL requirements} \\
    \midrule
    Advanced Arithmetic \& Math &
      Modern AI/DSP chips depend on rich numeric kernels; original is dominated by add/sub tasks. &
      Multi-operator datapaths, corner-case handling (overflow/saturation), and defined latency. \\

    Standard Communication Protocols &
      Real chips communicate via standardized serial protocols; absent in original set. &
      Framing/encoding, CRC/parity, start/stop conditions, and valid-ready handshakes. \\

    Memory, CDC \& Flow Control &
      Buffering is ubiquitous; single-clock primitives are insufficient for system robustness. &
      FIFO logic, full/empty semantics, backpressure, and safe control signaling. \\

    DSP \& Image Processing &
      Accelerators frequently embed DSP kernels; original tasks ignore structured data flow. &
      Windowed/streaming computation, pipeline alignment, and boundary handling. \\

    Advanced Timing \& System Control &
      Timing/latency errors are common tape-out bugs; benchmarks need multi-cycle complexity. &
      Multi-stage pipelines, coordinated enables, complex FSM + datapath interaction. \\
    \bottomrule
  \end{tabular}
\end{table}

\subsection{Grading Scheme}
\label{appendix:grading_scheme}

To enable a more fine-grained analysis, we stratify the 209 problems into Easy/Medium/Hard using a rule-based structural complexity score computed from the canonical RTL implementation. The complexity of hardware description language (HDL) code can be systematically decomposed into three orthogonal dimensions: (i) structural complexity, which is measured by the number of lines of code (LOC) to reflect the scale and modularity of the design; (ii) data complexity, usually reflected by the bit width for processing data to indicate numerical precision and storage requirements; and (iii) temporal complexity, which is characterized by the diversity of temporal structures and nesting depth to represent the complexity of temporal logic such as clock domains, state machines, and pipelines. 

These three dimensions jointly determine the difficulty of understanding, implementing, and verifying the design. During the construction process, these three dimensions are scored, with structural complexity mapped to the macroscopic indicator of lines of code (LOC); data complexity is mainly represented by the maximum bit width (Max Width); and temporal complexity is jointly characterized by the complexity of assignment statements (\#assign), the number of blocks (\#always), and conditional branches (\#case) among other syntactic elements.

Therefore, the complexity score $S$ is the sum of sub-scores derived from five metrics:
\begin{equation}
  S \;=\; s_{\mathrm{loc}} + s_{\mathrm{assign}} + s_{\mathrm{always}} + s_{\mathrm{case}} + s_{\mathrm{width}}
\end{equation}

\noindent The scoring rules are detailed in \autoref{tab:grading_rules}. We map the total score $S$ to difficulty labels:
\begin{equation*}
  \text{\colorbox{easybg}{\textbf{Easy}}: } S \le 1 \quad\quad
  \text{\colorbox{medbg}{\textbf{Medium}}: } 2 \le S \le 3 \quad\quad
  \text{\colorbox{hardbg}{\textbf{Hard}}: } S \ge 4.
\end{equation*}

\begin{table}[h]
  \centering
  \caption{Scoring rules based on the structural complexity of designs.}
  \label{tab:grading_rules}
  \appendixtablestyle
  \rowcolors{2}{verylightgray}{white}
  \begin{tabular}{@{}l l c@{}}
    \toprule
    \textbf{Metric} & \textbf{Thresholds} & \textbf{Points} \\
    \midrule
    \textbf{LOC} & $\le 10$ / $11\text{--}30$ / $31\text{--}60$ / $>60$ & $0$ / $1$ / $2$ / $3$ \\
    \textbf{\#Assign} & $\le 1$ / $2\text{--}4$ / $\ge 5$ & $0$ / $1$ / $2$ \\
    \textbf{\#Always} & $0$ / $1$ / $2$ / $\ge 3$ & $0$ / $1$ / $2$ / $3$ \\
    \textbf{\#Case} & $0$ / $1$ / $2$ / $\ge 3$ & $0$ / $1$ / $2$ / $3$ \\
    \textbf{Max Width} & $\le 32$ / $33\text{--}128$ / $>128$ & $0$ / $1$ / $2$ \\
    \bottomrule
  \end{tabular}
\end{table}

\paragraph{Task Lists.}
Applying this grading scheme, we obtain an initial difficulty label for each problem. We then manually reviewed the resulting split to ensure the final annotations are consistent with human-perceived implementation and verification difficulty. The final difficulty lists are:

\begin{tcolorbox}[
    colback=easybg, 
    colframe=easyframe, 
    title=\textbf{Easy (51 tasks)}, 
    fonttitle=\sffamily\bfseries,
    boxrule=0.8pt,
    arc=2pt
]
\scriptsize\ttfamily
Prob001, Prob002, Prob003, Prob004, Prob005, Prob006, Prob007, Prob008, Prob009, Prob010, Prob011, Prob012, Prob013, Prob014, Prob015, Prob016, Prob019, Prob020, Prob022, Prob023, Prob024, Prob025, Prob027, Prob028, Prob029, Prob031, Prob034, Prob042, Prob043, Prob050, Prob055, Prob056, Prob057, Prob062, Prob083, Prob090, Prob098, Prob101, Prob102, Prob103, Prob131, Prob169, Prob188, Prob191, Prob194, Prob195, Prob197, Prob201, Prob202, Prob203, Prob207.
\end{tcolorbox}

\begin{tcolorbox}[
    colback=medbg, 
    colframe=medframe, 
    title=\textbf{Medium (91 tasks)}, 
    fonttitle=\sffamily\bfseries,
    boxrule=0.8pt,
    arc=2pt
]
\scriptsize\ttfamily
Prob017, Prob018, Prob021, Prob026, Prob030, Prob032, Prob033, Prob035, Prob036, Prob037, Prob038, Prob039, Prob040, Prob041, Prob044, Prob045, Prob046, Prob047, Prob048, Prob049, Prob051, Prob052, Prob053, Prob054, Prob059, Prob060, Prob061, Prob063, Prob064, Prob065, Prob066, Prob067, Prob068, Prob069, Prob070, Prob071, Prob072, Prob073, Prob074, Prob075, Prob076, Prob077, Prob080, Prob081, Prob082, Prob084, Prob085, Prob091, Prob092, Prob093, Prob097, Prob099, Prob100, Prob104, Prob105, Prob106, Prob112, Prob113, Prob114, Prob116, Prob117, Prob118, Prob122, Prob123, Prob125, Prob126, Prob130, Prob132, Prob135, Prob141, Prob145, Prob163, Prob165, Prob168, Prob170, Prob171, Prob176, Prob178, Prob182, Prob183, Prob184, Prob186, Prob190, Prob193, Prob196, Prob198, Prob199, Prob200, Prob205, Prob208, Prob209.
\end{tcolorbox}

\begin{tcolorbox}[
    colback=hardbg, 
    colframe=hardframe, 
    title=\textbf{Hard (67 tasks)}, 
    fonttitle=\sffamily\bfseries,
    boxrule=0.8pt,
    arc=2pt
]
\scriptsize\ttfamily
Prob058, Prob078, Prob079, Prob086, Prob087, Prob088, Prob089, Prob094, Prob095, Prob096, Prob107, Prob108, Prob109, Prob110, Prob111, Prob115, Prob119, Prob120, Prob121, Prob124, Prob127, Prob128, Prob129, Prob133, Prob134, Prob136, Prob137, Prob138, Prob139, Prob140, Prob142, Prob143, Prob144, Prob146, Prob147, Prob148, Prob149, Prob150, Prob151, Prob152, Prob153, Prob154, Prob155, Prob156, Prob157, Prob158, Prob159, Prob160, Prob161, Prob162, Prob164, Prob166, Prob167, Prob172, Prob173, Prob174, Prob175, Prob177, Prob179, Prob180, Prob181, Prob185, Prob187, Prob189, Prob192, Prob204, Prob206.
\end{tcolorbox}

\subsection{Issues with Prob099}

During the construction and revision of our \textsc{VerilogEval-v2-EXT} Benchmark, we found an issue with the original Prob099 design from VerilogEval-v2 under our Verilator environment. Specifically, when we attempted to compile and run the Prob099 harness under Verilator, the build failed during elaboration with fatal named-port connection errors, so the simulation never started and no functional comparison could be performed. In the Verilator log, the failure presents as repeated \texttt{PINNOTFOUND} diagnostics indicating that the testbench tries to connect ports \texttt{Y2} and \texttt{Y4} on both the reference instance and the DUT instance, even though those formal port names do not exist on the corresponding module definitions. A representative excerpt is:
\begin{codebox}{bash}
%Error-PINNOTFOUND: ...Prob099_m2014_q6c_test.sv:74:4: Pin not found: 'Y2'
%Error-PINNOTFOUND: ...Prob099_m2014_q6c_test.sv:75:4: Pin not found: 'Y4'
%Error-PINNOTFOUND: ...Prob099_m2014_q6c_test.sv:80:4: Pin not found: 'Y2'
%Error-PINNOTFOUND: ...Prob099_m2014_q6c_test.sv:81:4: Pin not found: 'Y4'
%Error: Exiting due to 4 error(s)
\end{codebox}

The root cause is an interface mismatch between the testbench and the provided reference/DUT module definitions. The reference module \texttt{RefModule} (and the auto-generated \texttt{TopModule} wrapper used for validation) exports outputs named \texttt{Y1} and \texttt{Y3}, but the testbench file instantiates both modules as if they exported \texttt{Y2} and \texttt{Y4}. Verilator resolves named port connections strictly: when it sees an instantiation such as \texttt{.Y2(Y2\_ref)}, it searches the callee's port list for a formal named \texttt{Y2}; if no such formal exists, it raises \texttt{PINNOTFOUND} and terminates. Conceptually, this is consistent with a naming/offset confusion in the problem statement text: the prompt mentions ``\texttt{Y2} and \texttt{Y4} corresponding to \texttt{y[1]} and \texttt{y[3]},'' while the actual reference implementation uses the convention ``\texttt{Y1} corresponds to the next-state signal for \texttt{y[1]} and \texttt{Y3} corresponds to the next-state signal for \texttt{y[3]}.'' In other words, the authoritative artifacts (reference module + wrapper) define \texttt{Y1}/\texttt{Y3}, but the testbench was written against the inconsistent \texttt{Y2}/\texttt{Y4} naming. 

To fix the issue, we made the testbench consistent with the reference/DUT interface and with the benchmark's intended observation points. We renamed the internal signals and updated the named port connections in both instantiations accordingly. We also aligned the testbench's declaration of the state vector to the canonical module interface by changing it from a 1-based range (\texttt{logic [6:1] y}) to the conventional 0-based range (\texttt{logic [5:0] y}) in both the stimulus generator and the top-level testbench. After these changes, Verilator is able to compile the testbench successfully and execute the simulation, and the DUT (when wrapped to mirror the reference) produces matching \texttt{Y1} and \texttt{Y3} outputs over the randomized one-hot stimulus set.


\section{Additional Implementation Details}
\label{appendix:verisure_impl}

This appendix documents additional implementation-level details of \textsc{Veri-Sure} that are omitted from the main paper for clarity, including the design contract schema, linting rules, static dependency graph construction for localization, trace alignment/slicing for temporal debugging, and formal miter generation for Boolean equivalence checking.

\subsection{Contract Schema}
\label{appendix:contract_schema}

As described in \autoref{sec:vev2ext}, each benchmark task provides a fixed module interface and an executable testbench. Nevertheless, the intent behind cycle-accurate behaviors (reset conventions, sampling edge, output latency, corner cases) is frequently under-specified in natural language and can drift across iterations or agent handoffs. \textsc{Veri-Sure} addresses this by introducing a design contract as a structured intermediate representation (IR) shared across agents.

\paragraph{Core Fields.}
In our implementation, the contract is a JSON object with five core top-level sections:

\begin{itemize}
  \item \texttt{module\_name}: (\texttt{string}) The top-level DUT module identifier.
  \item \texttt{io}: (\texttt{list}) An ordered list of port objects. Each port entry contains:
  \begin{itemize}
    \item \texttt{name}: (\texttt{string}) Signal identifier (SystemVerilog identifier constraints).
    \item \texttt{dir}: (\texttt{string}) One of \{\texttt{input}, \texttt{output}, \texttt{inout}\}.
    \item \texttt{width}: (\texttt{int}) Positive integer bit-width (scalars use \texttt{1}).
    \item \texttt{description}: (\texttt{string}) Natural-language semantic summary.
  \end{itemize}
  \item \texttt{clocking}: (\texttt{object}) Global clock/reset semantics:
  \begin{itemize}
    \item \texttt{is\_sequential}: (\texttt{bool}) Whether the design holds clocked state. For purely combinational designs it is \texttt{false} and the clock/reset entries are omitted.
    \item \texttt{clock.name}, \texttt{clock.edge}: Clock input port and its active edge (\texttt{posedge} or \texttt{negedge}); required when \texttt{is\_sequential} is \texttt{true}.
    \item \texttt{reset.name}, \texttt{reset.active}, \texttt{reset.type}: Reset input port, its polarity (\texttt{high} or \texttt{low}), and whether it is \texttt{synchronous} or \texttt{asynchronous}; omitted when the design has no reset.
  \end{itemize}
  \item \texttt{timing}: (\texttt{object}) A mapping from each output to its expected latency, in cycles, relative to contract-defined sampling:
  \begin{itemize}
    \item \texttt{timing.outputs[signal].latency\_cycles}: (\texttt{int}) non-negative; \texttt{0} denotes combinational-by-contract.
  \end{itemize}
  \item \texttt{functional\_summary}: (\texttt{object}) A compact functional description used to ground verification, including Boolean/combinational obligations:
  \begin{itemize}
    \item Typically contains a high-level \texttt{overview} plus a machine-readable list of \texttt{rules} describing input/output relations and corner cases.
  \end{itemize}
\end{itemize}

\paragraph{Optional Fields.}
To support richer tasks, the implementation also allows optional keys (ignored if absent), such as \texttt{parameters} (typed module parameters) and \texttt{test\_plan} (directed stimulus/check intent).

\paragraph{Schema Usage.}
The same contract instance is injected into:
(i) the \textbf{Coder} prompt (to enforce interface/timing consistency),
(ii) the \textbf{Verifier} prompt (to generate a self-checking testbench consistent with the contract),
(iii) the \textbf{Debugger} prompt (as the reference that repairs must preserve, alongside the trace report),
(iv) the \textbf{Asserter} prompt (to generate temporal assertions for the clock/reset/latency obligations),
and (v) the \textbf{Boolean Proofer} (to extract combinational obligations for formal equivalence when applicable).

\subsection{Contract Linting}
\label{appendix:contract_linting}

To prevent malformed or hallucinated contracts from propagating into RTL code generation and verification, \textsc{Veri-Sure} applies a straightforward linter to every Architect-produced JSON before invoking downstream agents.

\paragraph{Linting Objectives.}
The linter reports each issue as an \emph{error} or a \emph{warning} and checks three classes of constraints:

\begin{enumerate}
  \item \textbf{Schema validity (structure and typing):}
  The contract must be a JSON object with a non-empty \texttt{module\_name} and a non-empty \texttt{io} list of port objects.
  The linter validates types (e.g., \texttt{width} is a positive integer, defaulting to \texttt{1} when omitted, and \texttt{latency\_cycles} is a non-negative integer)
  and rejects invalid enumerations (e.g., \texttt{dir} not in \{\texttt{input},\texttt{output},\texttt{inout}\}).

  \item \textbf{Signal consistency and referential integrity:}
  The linter ensures that port names are non-empty and unique, and that every output has a \texttt{timing} entry with a \texttt{latency\_cycles} value. A missing entry is reported as a warning, since the latency of that output becomes ambiguous.

  \item \textbf{Clock/reset integrity:}
  If \texttt{clocking.is\_sequential} is \texttt{true}, a missing \texttt{clocking.clock.name} is an error, and a clock or reset name that is not an input port is reported as a warning. Combinational designs and designs without a reset are therefore valid contracts.
\end{enumerate}

The linter also warns when the per-agent \texttt{guidance} block is missing. If the lint report contains errors, the Architect is re-invoked with the report to fix the contract, for up to two revision rounds before generation proceeds. Warnings are recorded in the lint report but do not trigger a revision.

\subsection{Static Dependency Graph Construction}
\label{appendix:dep_graph}

\textsc{Veri-Sure} localizes repairs using a static dependency slice over the RTL source (\autoref{sec:tracing_slicing_patching}). This slice is computed from a dependency graph built directly from the SystemVerilog AST, rather than fragile regex matching.

\paragraph{Parsing Frontend.}
We use \texttt{tree-sitter-verilog} to parse the generated RTL into an AST and extract semantic units. This provides stable node spans (start/end line ranges) required by our block-level patching mechanism.

\paragraph{Block Granularity.}
The RTL is decomposed into atomic \texttt{RtlBlock} units. A block is any semantic unit that can drive signals:
\begin{itemize}
  \item \textbf{Procedural blocks:} \texttt{always}/\texttt{always\_comb}/\texttt{always\_ff}/\texttt{always\_latch}.
  \item \textbf{Continuous assignments:} \texttt{assign} statements.
  \item \textbf{Submodule instances:} treated as blocks to account for hierarchical wiring when present.
\end{itemize}
Each block stores a unique identifier, its source span (line range), its raw text, and statically extracted read/write sets.

\paragraph{Read/Write Extraction.}
For each block $B$, we compute:
\begin{itemize}
  \item $W(B)$ (\textbf{write set}): all L-values written by the block (e.g., left-hand sides of \texttt{=} or \texttt{<=}, including bit/part selects mapped to their base identifiers).
  \item $R(B)$ (\textbf{read set}): all identifiers used as R-values in RHS expressions and control predicates (e.g., \texttt{if} conditions, \texttt{case} selectors), as well as explicit sensitivity list identifiers when present.
\end{itemize}
For module instances, $W(B)$ and $R(B)$ are inferred from the port connections when directionality is unambiguous; otherwise, the instance is conservatively treated as reading all connected signals, and writing those connected to nets whose names match the instance's output connections in the AST.

\paragraph{Driver Map and Dependency Relation.}
We build a global driver map:
\begin{equation}
  \mathcal{D}(s) \;=\; \{\, B \mid s \in W(B)\,\},
\end{equation}
which maps each signal $s$ to the set of blocks that may drive it.
A directed dependency edge exists from $B_i$ to $B_j$ iff $B_j$ reads a signal written by $B_i$:
\begin{equation}
  B_j \leftarrow B_i
  \quad \Leftrightarrow \quad
  W(B_i)\cap R(B_j)\neq\varnothing.
\end{equation}

\paragraph{Backward Dependency Slicing.}
Given a failing output set $\mathcal{Y}_{\mathrm{fail}}$ identified by simulation/trace analysis, we compute the suspect block set $\mathcal{B}_{\mathrm{sus}}$ using a bounded backward BFS:
\begin{align}
  \mathcal{B}_0 &= \bigcup_{y \in \mathcal{Y}_{\mathrm{fail}}} \mathcal{D}(y), \\
  \mathcal{B}_{d+1} &= \mathcal{B}_{d} \;\cup\;
  \bigcup_{B \in \mathcal{B}_d}
  \bigcup_{r \in R(B)} \mathcal{D}(r),
\end{align}
and finally $\mathcal{B}_{\mathrm{sus}} = \mathcal{B}_{d_{\max}}$.
We use a small fixed depth bound $d_{\max}{=}2$, so the slice spans at most three levels of driver blocks, to balance localization precision and completeness: deeper slices increase recall but also inflate the Debugger context window, which can reduce the precision of LLM-guided patches.

The resulting $\mathcal{B}_{\mathrm{sus}}$ is then used to enforce edit locality: the Debugger is only allowed to modify blocks within $\mathcal{B}_{\mathrm{sus}}$, while all other lines are preserved verbatim, reducing regressions and ``repair hallucinations''.

\subsection{Trace Alignment \& Slicing}
\label{appendix:trace_alignment_slicing}

Simulation feedback is often too coarse (pass/fail or a single mismatch) to support reliable multi-cycle repairs. \textsc{Veri-Sure} therefore converts raw simulation artifacts into an LLM-consumable trace report that localizes failures in both time and signal space.

\paragraph{Failure Localization.}
The Verilator harness captures stdout/stderr and parses for the earliest failure event, including assertion failures emitted by the Asserter or explicit mismatch diagnostics (e.g., ``Mismatch at time \dots''). This produces the first failing timestamp $t_f$ (or equivalently a failure cycle index under the contract-defined sampling scheme).

\paragraph{Windowed Waveform Extraction.}
We extract a short window $\mathcal{W}(t_f)$ ending at the failure, consisting of the last $L{=}10$ clock edges (posedge and negedge, i.e., five clock cycles) at or before $t_f$.
Using a VCD parser, we sample signal values at every edge in the window, label each sample with its edge type, and record:
\begin{enumerate}
  \item the expected and observed values of the failing output signals $\mathcal{Y}_{\mathrm{fail}}$,
  \item all top-level inputs (for context),
  \item up to 20 internal signals identified by the static slice (signals appearing in $R(B)$ or $W(B)$ for $B\in\mathcal{B}_{\mathrm{sus}}$, excluding top-level ports).
\end{enumerate}

\paragraph{Alignment Diagnosis for Off-by-One Timing Bugs.}
Pipeline and handshake designs frequently fail due to latency mismatches or sampling-edge errors.
To provide actionable timing hints, the tracing mechanism performs a lightweight alignment check on each failing output over a longer history than $\mathcal{W}(t_f)$. Let $e_k$ and $a_k$ denote the expected and observed values of the output at the $k$-th clock edge (posedge or negedge), taken from the last $N{=}200$ edges up to $t_f$. For a shift $\delta$, measured in clock edges, we compute the match rate
\begin{equation}
  \mathrm{match}(\delta) \;=\;
  \frac{1}{|K_\delta|}
  \sum_{k \in K_\delta}
  \mathbb{I}\!\left[a_k = e_{k-\delta}\right],
  \qquad \delta \in \{-1, 0, +1\},
\end{equation}
where $K_\delta$ is the set of indices $k$ for which both $a_k$ and $e_{k-\delta}$ are available. Under this convention, $\delta = +1$ means that the DUT output lags the expected output by one edge, and $\delta = -1$ means that it leads. We also compute $\mathrm{match}(0)$ separately over posedge and over negedge samples to expose wrong-edge updates. The report includes these rates together with the best-matching shift; for example, a best match at $\delta = +1$ indicates that the output appears one edge late.
This signal is complementary to assertion failures generated by the Asserter and is particularly effective for diagnosing systematic latency shifts that may not be obvious from single-cycle mismatches.

\paragraph{Trace Report Format.}
The final trace report includes:
(i) the failure cycle/time, (ii) the set of failing outputs, (iii) the best alignment hint (if any), and (iv) the windowed trace slice.
This report is injected into the Debugger prompt together with the localized suspect blocks $\mathcal{B}_{\mathrm{sus}}$, enabling the Debugger to propose minimal edits grounded in concrete temporal evidence.

\subsection{Formal Miter Construction}
\label{appendix:formal_miter}

The Boolean Proofer targets purely combinational logic by proving equivalence between (i) the DUT implementation and (ii) a compact specification model synthesized from the contract. This is implemented using an automatically generated miter.

\paragraph{Selecting Targets of Proof.}
From the contract timing map, we identify candidate combinational outputs:
\begin{equation}
  \mathcal{Y}_{\mathrm{comb}} \;=\; \{\, y \in \text{Outputs} \mid \texttt{latency\_cycles}(y)=0 \,\}.
\end{equation}
To avoid unsound proofs on stateful logic, we further filter $\mathcal{Y}_{\mathrm{comb}}$ by dependency analysis: any output driven by an \texttt{always\_ff} (or a clocked \texttt{always}) in the current RTL is excluded from combinational proving. This conservative rule ensures the miter remains purely combinational.

\paragraph{Model Synthesis and Wrapping.}
We compile \texttt{functional\_summary} into a synthesizable SystemVerilog specification module that computes each $y_{\mathrm{spec}}$ in $\mathcal{Y}_{\mathrm{comb}}$ from the shared inputs $\mathbf{x}$. For simple Boolean expressions, this is emitted as \texttt{assign} statements; for piecewise logic, it is emitted as \texttt{always\_comb} with \texttt{if}/\texttt{case}. 

We then generate a wrapper that instantiates the DUT and the specification module in parallel, ties their inputs, and asserts equality on selected outputs. A simplified sketch is shown below:

\begin{titlecodebox}{verilog}{Example}
module Miter(/* shared inputs */);
  // shared inputs: declared here
  // dut/spec outputs
  logic [W-1:0] y_dut, y_spec;

  dut_top dut (/* .a(a), .b(b), ... */, .y(y_dut));
  SpecModule spec (/* .a(a), .b(b), ... */, .y(y_spec));

  always @* begin
    assert (y_dut == y_spec);
  end
endmodule
\end{titlecodebox}

\paragraph{Running SymbiYosys.}
The DUT, specification module, and Miter are passed to SymbiYosys~\citep{yosyshq_symbiyosys} using an SMT-based proving flow (Z3~\citep{demoura2008z3} backend in our setup). Since the miter is combinational, the proof reduces to a SAT/SMT validity query over $\forall \mathbf{x}$.
If the proof fails, \texttt{sby} produces a counterexample assignment $\mathbf{x}^\star$; we parse this model and translate it into a concrete input vector (and, when applicable, a directed stimulus snippet) that is fed back to the Debugger. This provides a precise, minimal witness of the Boolean mismatch and helps disambiguate pure logic faults from timing/latency faults already handled by the trace and assertion branches.

\section{Additional Experimental Results}
\label{appendix:experiments}

\subsection{Failure Mode Analysis}
\label{appendix:failure-analysis}

We manually inspected the 14 tasks that did not pass after the full \textsc{Veri-Sure} loop. One task was a tooling artifact, leaving 13 true failures. The breakdown in \autoref{tab:failure-modes} shows that the remaining bottleneck is not syntax generation, but long-horizon temporal reasoning and numerically delicate combinational semantics.

\begin{table}[h]
\centering
\caption{Failure mode breakdown for non-passing \textsc{Veri-Sure} outputs on \textsc{VerilogEval-v2-EXT}. Percentages are computed over the 13 true failures, excluding one tooling artifact.}
\label{tab:failure-modes}
\appendixtablestyle
\rowcolors{2}{verylightgray}{white}
\begin{tabular}{@{}L{0.26\textwidth}cL{0.53\textwidth}@{}}
\toprule
\textbf{Failure type} & \textbf{Count} & \textbf{Typical root cause} \\
\midrule
Sequential / timing & 10 (76.9\%) & Wrong FSM transitions, off-by-one latency, reset semantics, wrong-edge updates, or pulse alignment errors. Examples include cases where assertion traces localize a wrong-edge update or a one-cycle pulse shift, but the repair budget is exhausted before a fully correct patch is found. \\
Hard combinational arithmetic & 3 (23.1\%) & Numerically delicate arithmetic, especially floating-point normalization, subnormal handling, rounding, or width-sensitive corner cases. These failures usually require more precise numerical specifications than the contract's functional summary provides. \\
\midrule
Tooling artifact & 1 & Non-semantic failure due to checker/tool integration rather than an RTL behavior error; excluded from the true-failure percentages above. \\
\bottomrule
\end{tabular}
\end{table}

This distribution explains the gap between high syntax pass rates and lower functional pass rates on the hardest tasks. Once syntax and simple combinational logic are solved, the dominant failures arise from temporal corner cases that require consistent reasoning across multiple cycles, or from arithmetic specifications whose corner cases are difficult to infer from natural language alone.

\subsection{Design Contract Audit}
\label{appendix:contract-audit}

Since all downstream agents condition on the Architect's contract, we audited whether it faithfully captures the natural-language specification. We manually compared each of the 209 contracts produced in the main GPT-5.2 run against its specification and recorded functional contradictions, constraints not supported by the specification (hallucinations), and interface errors. \autoref{tab:contract-audit} summarizes the results: 12 contracts (5.7\%) contain at least one error.

\begin{table}[h]
\centering
\caption{Manual audit of the 209 design contracts from the main GPT-5.2 run. The 12 erroneous contracts contain 13 errors in total.}
\label{tab:contract-audit}
\appendixtablestyle
\rowcolors{2}{verylightgray}{white}
\begin{tabular}{@{}L{0.26\textwidth}cL{0.53\textwidth}@{}}
\toprule
\textbf{Error type} & \textbf{Count} & \textbf{Typical cases} \\
\midrule
Functional contradiction & 10 & K-map derivation, waveform reproduction, and FSM output semantics. \\
Hallucinated constraint & 3 & Requirements added by the Architect that the specification does not state. \\
Interface error & 0 & -- \\
\midrule
Erroneous contracts & 12 (5.7\%) & 3 overlap with the 13 true failures in \autoref{tab:failure-modes}; the other 9 tasks were solved. \\
\bottomrule
\end{tabular}
\end{table}

Only 3 of the 12 affected tasks are among the true failures. For the other 9, the Debugger overrides the faulty contract because the trace report is computed from simulation against the golden testbench rather than from the contract. To test whether the 3 overlapping failures are caused by their contract errors, we re-ran them with the contract removed; all 3 still fail, indicating that these tasks are intrinsically hard rather than bottlenecked by the contract. Together with the ablation in \autoref{tab:ablation_singlecol} (w/o Design Contract, $-2.9$ pp), this suggests that the contract is a net gain and that the trace-driven debugging loop limits the impact of its occasional errors.

\subsection{Reliability of LLM-Generated Formal Signals}
\label{appendix:formal-signal-reliability}

The Asserter and Boolean Proofer generate formal artifacts from an LLM-derived structured design specification, so \textsc{Veri-Sure} uses them as diagnostic hints for repair and keeps the final acceptance gated by the golden testbench. Erroneous or incomplete formal hints can consume repair budget, but they cannot by themselves certify an incorrect design as correct. We inspected formal-signal behavior on non-passing tasks and summarize the main observations in \autoref{tab:formal-signal-reliability}.

\begin{table}[h]
\centering
\caption{Reliability analysis of LLM-generated formal artifacts. Formal signals improve diagnosis on many tasks, but are not used as acceptance oracles.}
\label{tab:formal-signal-reliability}
\appendixtablestyle
\rowcolors{2}{verylightgray}{white}
\begin{tabular}{@{}L{0.20\textwidth}L{0.35\textwidth}L{0.37\textwidth}@{}}
\toprule
\textbf{Artifact} & \textbf{Useful behavior} & \textbf{Observed limitation} \\
\midrule
Asserter & Produces concrete timing diagnostics, including wrong-edge updates, reset mismatches, and off-by-one latency violations. These diagnostics often identify the first failing cycle and implicated signals. & Can be incomplete when the generated monitor omits the relevant temporal condition. For example, an SPI release-edge failure can produce simulation mismatches while the generated assertion set reports no trigger. \\
Boolean Proofer & Provides counterexamples for latency-0 combinational obligations after dependency filtering, helping separate pure Boolean faults from sequential bugs. & Can fail to produce useful guidance when the generated reference model uses unsupported constructs or oversimplifies complex numerical semantics, such as floating-point normalization and subnormal handling. \\
Acceptance gate & Prevents false acceptance because repaired RTL must still pass the executable golden testbench. & Does not guarantee exhaustive silicon-level correctness; formal artifacts are auxiliary diagnostics rather than complete specifications. \\
\bottomrule
\end{tabular}
\end{table}

\subsection{Repair Statistics}
\label{appendix:repair-stats}

To quantify how much the debugging loop changes a design, we collected repair statistics from the saved results of the main GPT-5.2 run. Of the 209 tasks, 169 (80.9\%) pass without any repair. Averaged over all tasks, \textsc{Veri-Sure} applies 1.14 fixes per task, or 5.97 fixes per task that needed repair. A localized patch modifies 7.9 lines (292 characters) on average, whereas regenerating the whole file at every fix rewrites 68.8 lines (1{,}944 characters). Each patch therefore touches only about 10\% of the content a regeneration rewrites, and the rest of the file is preserved verbatim. Locality also improves accuracy: replacing localized patching with whole-file regeneration lowers overall functional Pass@1 from 93.30\% to 89.95\% (w/o Patching only in \autoref{tab:ablation_singlecol}).

\subsection{Cost and Token Analysis}
\label{appendix:cost-analysis}

\autoref{tab:cost-analysis} reports per-task token usage and API cost for GPT-5.2-backed agentic methods on \textsc{VerilogEval-v2-EXT}. All rows use the same backbone to isolate framework-level differences. The cached column reflects prompt-cache discounts for repeated system instructions, tool descriptions, and stable context prefixes. We report both raw and cached costs to separate worst-case API cost from realistic repeated-run deployment cost.

\begin{table}[h]
\centering
\caption{Per-task token and API cost analysis with GPT-5.2 as the backbone. Costs are averaged over tasks.}
\label{tab:cost-analysis}
\appendixtablestyle
\rowcolors{2}{verylightgray}{white}
\begin{tabular}{@{}lrrrr@{}}
\toprule
\textbf{Method} & \textbf{Input tok.} & \textbf{Output tok.} & \textbf{\$/task} & \textbf{\$/task (cache)} \\
\midrule
MAGE (10 samples) & 16{,}677 & 5{,}508 & 0.806 & 0.798 \\
VerilogCoder & 25{,}872 & 11{,}718 & 0.209 & 0.197 \\
\textsc{Veri-Sure} & 62{,}203 & 3{,}653 & 0.160 & 0.131 \\
\bottomrule
\end{tabular}
\end{table}

\textsc{Veri-Sure} uses more input tokens because it carries structured diagnostic context, traces, and formal hints across the repair loop. However, localized patching substantially reduces output tokens by avoiding whole-file regeneration. Since output tokens are more expensive than input tokens in the measured API pricing, \textsc{Veri-Sure} is cheaper than other multi-agent frameworks while solving more tasks in the main evaluation.

\section{Case Studies}
\label{appendix:case_study}

\subsection{Case Study 1: Boolean Proofer for an 8-bit Signed Multiplier}
\label{appendix:case_study_multiplier}

This case study evaluates the effectiveness of the Boolean Proofer on an 8-bit parallel multiplier (\texttt{Prob157}) designed using two's-complement arithmetic. The objective is to implement a purely combinational multiplier that calculates the product of two 8-bit signed integers, $a$ and $b$, by summing eight sign-extended and shifted 16-bit partial products. This corresponds to the first case in~\autoref{fig:case_study}.

\paragraph{Problem Description.}
The initial candidate implementation failed simulation, so the Debugger invoked the Boolean Proofer to check it against the contract-derived reference model. During Boolean equivalence checking, the proofer identified a fundamental mismatch between the implementation logic and the mathematical specification of signed multiplication.

\paragraph{Formal Specification and Counterexample.}
The Boolean Proofer identified that the implementation treated the multiplier $b$ as an unsigned value, failing to account for the negative weight of the Most Significant Bit (MSB). The logical divergence is summarized below:

\begin{itemize}
    \item \textbf{Erroneous Implementation Logic:}
    \[P = \left( -a[7] \cdot 2^7 + \sum_{i=0}^6 a[i] \cdot 2^i \right) \times \left( \sum_{j=0}^7 b[j] \cdot 2^j \right)\]
    
    \item \textbf{Correct Specification Logic:}
    \[P = \left( -a[7] \cdot 2^7 + \sum_{i=0}^6 a[i] \cdot 2^i \right) \times \left( -b[7] \cdot 2^7 + \sum_{j=0}^6 b[j] \cdot 2^j \right)\]
\end{itemize}

\paragraph{Root Cause Identification.}
The error stemmed from a common misconception in signed arithmetic. In a two's-complement multiplier, the MSB of the multiplier, $b[7]$, carries a negative weight ($-2^7$).

\begin{enumerate}
    \item \textbf{Unsigned Weight Assumption}: The original logic treated all bits of $b$ as positive weights, effectively calculating $A \times B_{\mathrm{unsigned}}$.
    \item \textbf{Partial Product Sign Error}: For a correct signed product, the 8th partial product must be subtracted (or added as a two's-complement negation) to account for the negative weight:
    \[\mathrm{Product} = \sum_{i=0}^{6} (b[i] \cdot (a \ll i)) - (b[7] \cdot (a \ll 7))\]
\end{enumerate}

\paragraph{Code Patching.}
Following the feedback from the Boolean Proofer, the Verilog implementation was refined. The logic for generating the 8th partial product was adjusted to handle the negative weight of the sign bit as shown in \autoref{lst:multiplier_comparison}.

\begin{listing}[h!]
\caption{Code correction for Case Study 1.}
\label{lst:multiplier_comparison}
\begin{listingframe}
\begin{lstlisting}[style=diffstyle]
--- buggy_multiplier.sv
+++ corrected_multiplier.sv
@@ -1,8 +1,6 @@
-// BUGGY: All bits treated as positive weights (unsigned multiplication)
-a_ext = { {8{a[7]}}, a };                 // Sign-extend a to 16 bits
-for (i = 0; i < 8; i = i + 1) begin       // Loop over all 8 bits
-    if (b[i])                             // If bit is set
-        partial[i] = a_ext <<< i;         // Add positive shifted value
-    else
-        partial[i] = 16'sd0;              // Otherwise add zero
-end
+// CORRECTED: b[7] treated as negative weight (signed multiplication)
+a_ext = { {8{a[7]}}, a };                     // Sign-extend a to 16 bits
+for (i = 0; i < 7; i = i + 1) begin           // Bits 0-6: positive weights
+    partial[i] = b[i] ? (a_ext <<< i) : 16'sd0;
+end
+partial[7] = b[7] ? -(a_ext <<< 7) : 16'sd0;  // Bit 7: negative weight
\end{lstlisting}
\end{listingframe}
\end{listing}

The patch changes only the eighth partial product, which is now subtracted when $b[7]=1$. Re-running the Boolean Proofer on the patched RTL found no counterexample against the contract-derived reference model, and the design passed the golden testbench.

\subsection{Case Study 2: Assertion Violation in a Global History Register}
\label{appendix:case_study_assertion}

This case study examines \texttt{Prob118}, a 32-bit global branch-history register \texttt{GH} with an active-high asynchronous reset \texttt{areset}. On a misprediction (\texttt{train\_mispredicted}), the register is reloaded with \texttt{\{train\_history[30:0], train\_taken\}}; otherwise, a valid prediction (\texttt{predict\_valid}) shifts \texttt{predict\_taken} into it. The register value is exposed through the output \texttt{predict\_history}. This corresponds to the second case in~\autoref{fig:case_study}.

\paragraph{Problem Description.}
The contract specifies a rising-edge clock and an asynchronous active-high reset, so \texttt{predict\_history} should change only after a rising clock edge or when \texttt{areset} is asserted. From these clocking obligations, the Asserter generated a \texttt{NO\_NEGEDGE\_UPDATE} monitor intended to catch updates on the wrong clock edge. Like all Asserter outputs, the monitor is bound to the DUT and evaluated during Verilator simulation rather than by a formal solver.

\paragraph{Generated Assertion and Violation Report.}
\autoref{lst:assertion_property} shows the generated monitor. At every falling clock edge outside reset, it compares \texttt{predict\_history} with the value recorded at the previous falling edge and logs a violation if the two differ.

\begin{listing}[h!]
\caption{Asserter-generated monitor for the negative-edge stability check.}
\label{lst:assertion_property}
\begin{listingframe}
\begin{lstlisting}[style=diffstyle]
always @(negedge clk) begin
    ph_prev_negedge <= predict_history;
    if (!areset) begin
        if (predict_history !== ph_prev_negedge) begin
            asserter_log("NO_NEGEDGE_UPDATE",
                $sformatf("predict_history changed on negedge clk (wrong edge?). prev=%h now=%h",
                          ph_prev_negedge, predict_history));
        end
    end
end
\end{lstlisting}
\end{listingframe}
\end{listing}

During simulation, the monitor reported its first violation at $t=20$, where \texttt{predict\_history} changed from \texttt{0x00000000} to \texttt{0x00000001}. The report passes the violated property, the failure time, and the implicated signal values to the Debugger as a formal hint.

\paragraph{Root Cause Identification.}
Starting from \texttt{predict\_history}, the static slice contains two blocks: the continuous assignment that drives the output and the clocked \texttt{always} block that updates \texttt{GH}. In the original RTL, the output is a direct copy of the register (\texttt{assign predict\_history = GH;}), so its reset behavior is only implied by the reset branch of the clocked block. Guided by the Asserter's wrong-edge/reset hint, the Debugger kept the clocked update logic unchanged and targeted the output assignment.

\paragraph{Code Patching.}
The patch, shown in~\autoref{lst:assertion_fix_comparison}, rewrites only the output assignment so that \texttt{predict\_history} is forced to zero while \texttt{areset} is asserted; the clocked block is preserved verbatim.

\begin{listing}[h!]
\caption{Code correction for Case Study 2.}
\label{lst:assertion_fix_comparison}
\begin{listingframe}
\begin{lstlisting}[style=diffstyle]
--- buggy_register.sv
+++ corrected_register.sv
@@ -1,14 +1,14 @@
     always @(posedge clk or posedge areset) begin
         if (areset) begin
             GH <= 32'h0;
         end else begin
             if (train_mispredicted)
                 GH <= {train_history[30:0], train_taken};
             else if (predict_valid)
                 GH <= {GH[30:0], predict_taken};
             // else: hold value implicitly
         end
     end

-    assign predict_history = GH;
+    assign predict_history = areset ? 32'h0 : GH;
 endmodule
\end{lstlisting}
\end{listingframe}
\end{listing}

The patched module was re-simulated and passed the golden testbench. Because the edit touches a single line, the rest of the design is preserved exactly, which is the behavior the localized patching mechanism is designed to enforce.

\subsection{Case Study 3: Automated Repair of Shift Direction Mismatch}
\label{appendix:case_study_shift}

This case study illustrates a common logical discrepancy between intended RTL behavior and the actual implementation within a multi-functional sequential circuit in \texttt{Prob063}. The target design is a 4-bit register integrating two primary operations: (i) a left-shift with serial input \texttt{data}, and (ii) a modulo-16 decrement counter.

\paragraph{Problem Description.}
A functional mismatch was identified when both control signals, \texttt{shift\_ena} and \texttt{count\_ena}, were asserted. Per the specification, \texttt{shift\_ena} has higher priority; therefore, the register must perform a left shift regardless of \texttt{count\_ena}.

\paragraph{Trace Analysis \& Localization.}
The failure was captured at simulation timestamp $t=370$\,ns. Inputs are sampled at \texttt{negedge clk} and the register updates at the subsequent \texttt{posedge clk}. The trace log (summarized in \autoref{tab:trace_log}) records the pre-state $q(t)$ at the sampling edge and the observed/expected next-state values at the following update edge.

As shown in the trace, the design is in state \texttt{q(t)=4'b1100} with \texttt{shift\_ena=1} and \texttt{data=0}. Under the required left-shift specification, the next state must be
\texttt{\{q[2:0], data\}} $=$ \texttt{\{100, 0\}} $=$ \texttt{4'b1000}.
However, the DUT produces \texttt{4'b0110}, which corresponds to a right-shift style concatenation \texttt{\{data, q[3:1]\}}.
The automated debugger localized the failure to the shift transition logic via trace-driven temporal diagnosis, highlighting a bit-ordering error in the concatenation.

\begin{table}[htbp]
\centering
\caption{Failing transaction captured at $t=370$\,ns (inputs sampled at \texttt{negedge clk}; next state observed at the following \texttt{posedge clk}).}
\label{tab:trace_log}
\appendixtablestyle
\rowcolors{2}{verylightgray}{white}
\begin{tabular}{@{}lll@{}}
\toprule
\textbf{Signal / State} & \textbf{Value} & \textbf{Description} \\
\midrule
\texttt{clk} & 0 (\texttt{negedge}) & Sampling point for input vectors \\
\texttt{shift\_ena} & 1 & Shift operation enabled (higher priority) \\
\texttt{count\_ena} & 1 & Decrement enabled (masked by \texttt{shift\_ena}) \\
\texttt{data} & 0 & Serial input bit for shifting \\
\texttt{q(t)} & 1100 & Register state before update \\
\midrule
$q_{\mathrm{ref}}(t{+}1)$ & 1000 & Expected next state: \texttt{\{q[2:0], data\}} \\
$q_{\mathrm{dut}}(t{+}1)$ & 0110 & Observed next state: \texttt{\{data, q[3:1]\}} \\
\bottomrule
\end{tabular}
\end{table}

\paragraph{Root Cause Identification.}
The localization tool identified the suspect logic within the \texttt{always\_ff} block. Inspection of the RTL source confirms a directional implementation error: the design specification requires a left shift, but the implementation used a right-shift concatenation, placing \texttt{data} at the MSB rather than the LSB.

\paragraph{Code Patching.}
To resolve the discrepancy, the framework applied a localized patch that re-orders the concatenation to match the left-shift specification (placing \texttt{data} at the LSB), as shown in \autoref{lst:shift_diff}.

\begin{listing}[htbp]
\caption{Localized code correction for the shift-direction bug.}
\label{lst:shift_diff}
\begin{listingframe}
\begin{lstlisting}[style=diffstyle]
--- buggy_shift.v
+++ corrected_shift.v
@@ -2,7 +2,7 @@
 always_ff @(posedge clk) begin
     if (shift_ena) begin
         // Shift operation
-        q <= {data, q[3:1]};      // Right-shift (BUG)
+        q <= {q[2:0], data};      // Left-shift (FIXED)
     end else if (count_ena) begin
         q <= q - 4'd1;
     end else begin
\end{lstlisting}
\end{listingframe}
\end{listing}

Subsequent re-simulation confirmed that the revised logic matches the golden model for all exercised traces, including cases where \texttt{shift\_ena} and \texttt{count\_ena} are simultaneously asserted. By using localized patching instead of full-file regeneration, the framework eliminated the mismatch while preserving the stability of the remaining logic.

\section{Detailed Framework Comparison}
\label{appendix:framework_comparison}

This appendix provides a comparison between \textsc{Veri-Sure} and state-of-the-art frameworks, analyzing architectural differences, capability gaps, and performance characteristics.

\autoref{tab:framework_architecture} provides a systematic comparison of core architectural features and capabilities across leading frameworks. The table uses the following notation: \good : Fully supported / High capability, \partialgood : Partially supported / Medium capability, \bad : Not supported / Low capability, \textbf{--} : Not applicable.

\begin{table*}[h!]
\centering
\caption{Comprehensive architectural and capability comparison of RTL generation frameworks.}
\label{tab:framework_architecture}
\small
\renewcommand{\arraystretch}{1.2}
\resizebox{\textwidth}{!}{%
\begin{tabular}{@{}p{0.22\textwidth}ccccc@{}}
\toprule
\textbf{Feature / Capability} & \textbf{\textsc{Veri-Sure}} & \textbf{MAGE} & \textbf{VerilogCoder} & \textbf{AutoChip} & \textbf{Single-Agent} \\
\midrule

\rowcolor{black!5}
\multicolumn{6}{l}{\textbf{Core Architecture}} \\

Core Philosophy & \textcolor{ForestGreen}{\textbf{Contract \& Verification}} & Sampling & Planning \& AST & Feedback & Feedback \\
Agents & \textcolor{ForestGreen}{\textbf{6 Specialized}} & 4 Specialized & 4 Specialized & 2 General & 1 General \\
Communication & \textcolor{ForestGreen}{\textbf{Structured Contract}} & Shared Context & Shared Context & Shared Context & Single Prompt \\

\rowcolor{black!5}
\multicolumn{6}{l}{\textbf{Test \& Verification}} \\

Testing Approach & \textcolor{ForestGreen}{\textbf{Hybrid (Sim+Formal)}} & Simulation-only & Simulation-only & Simulation-only & Simulation-only \\
Formal Verification & \good & \bad & \bad & \bad & \bad \\
Assertion Generation & \good & \bad & \bad & \bad & \bad \\
Testbench Generation & \good & \partialgood & \partialgood & \bad & \bad \\

\rowcolor{black!5}
\multicolumn{6}{l}{\textbf{Debugging \& Repair}} \\

Repair Strategy & \textcolor{ForestGreen}{\textbf{Patching}} & Whole-file & Whole-file & Whole-file & Whole-file \\
Error Localization & \textcolor{ForestGreen}{\textbf{Block-level}} & Module-level & Module-level & File-level & File-level\\

\rowcolor{black!5}
\multicolumn{6}{l}{\textbf{Performance \& Efficiency}} \\

Cost Efficiency & \textcolor{orange}{Medium} & \textcolor{red}{Low} & \textcolor{red}{Low} &\textcolor{ForestGreen}{\textbf{High}} & \textcolor{ForestGreen}{\textbf{High}}  \\
Iteration Efficiency & \textcolor{ForestGreen}{\textbf{High}} & \textcolor{orange}{Medium} & \textcolor{red}{Low} & \textcolor{red}{Low} & \textcolor{red}{Low} \\
Regression Safety & \textcolor{ForestGreen}{\textbf{High}} & \textcolor{orange}{Medium} & \textcolor{red}{Low} & \textcolor{red}{Low} & \textcolor{red}{Low} \\

\bottomrule
\end{tabular}%
}
\end{table*}

\section{Prompts of Agents}

The full prompts of all agents are available in our \href{https://github.com/xyjoey/Veri-Sure}{GitHub repository}.

\end{document}